\documentclass{aastex63}

\usepackage{textcomp}
\usepackage[normalem]{ulem}

\usepackage{amsmath}

\usepackage[utf8]{inputenc}

\usepackage{hyperref}
\usepackage{multirow}
\usepackage{amsmath}
\usepackage{graphicx} 

\shorttitle{Tracing H$\alpha$ Fibrils through Bayesian Deep Learning}
\shortauthors{Jiang et al.}

\begin{document}

\title{{\bf \large Tracing H$\alpha$ Fibrils through Bayesian Deep Learning}}

\author{Haodi Jiang}
\affiliation{Institute for Space Weather Sciences, New Jersey Institute of Technology, University Heights, Newark, NJ 07102-1982, USA;
hj78@njit.edu, wangj@njit.edu, haimin.wang@njit.edu}
\affiliation{Department of Computer Science, New Jersey Institute of Technology, University Heights, Newark, NJ 07102-1982, USA}

\author{Ju Jing}
\affiliation{Institute for Space Weather Sciences, New Jersey Institute of Technology, University Heights, Newark, NJ 07102-1982, USA;
hj78@njit.edu, wangj@njit.edu, haimin.wang@njit.edu}
\affiliation{Center for Solar-Terrestrial Research, New Jersey Institute of Technology, University Heights, Newark, NJ 07102-1982, USA}
\affiliation{Big Bear Solar Observatory, New Jersey Institute of Technology, 40386 North Shore Lane, Big Bear City, CA 92314-9672, USA}

\author{Jiasheng Wang}
\affiliation{Institute for Space Weather Sciences, New Jersey Institute of Technology, University Heights, Newark, NJ 07102-1982, USA;
hj78@njit.edu, wangj@njit.edu, haimin.wang@njit.edu}
\affiliation{Center for Solar-Terrestrial Research, New Jersey Institute of Technology, University Heights, Newark, NJ 07102-1982, USA}
\affiliation{Big Bear Solar Observatory, New Jersey Institute of Technology, 40386 North Shore Lane, Big Bear City, CA 92314-9672, USA}

\author{Chang Liu}
\affiliation{Institute for Space Weather Sciences, New Jersey Institute of Technology, University Heights, Newark, NJ 07102-1982, USA;
hj78@njit.edu, wangj@njit.edu, haimin.wang@njit.edu}
\affiliation{Center for Solar-Terrestrial Research, New Jersey Institute of Technology, University Heights, Newark, NJ 07102-1982, USA}
\affiliation{Big Bear Solar Observatory, New Jersey Institute of Technology, 40386 North Shore Lane, Big Bear City, CA 92314-9672, USA}

\author{Qin Li}
\affiliation{Institute for Space Weather Sciences, New Jersey Institute of Technology, University Heights, Newark, NJ 07102-1982, USA;
hj78@njit.edu, wangj@njit.edu, haimin.wang@njit.edu}
\affiliation{Center for Solar-Terrestrial Research, New Jersey Institute of Technology, University Heights, Newark, NJ 07102-1982, USA}
\affiliation{Big Bear Solar Observatory, New Jersey Institute of Technology, 40386 North Shore Lane, Big Bear City, CA 92314-9672, USA}

\author{Yan Xu}
\affiliation{Institute for Space Weather Sciences, New Jersey Institute of Technology, University Heights, Newark, NJ 07102-1982, USA;
hj78@njit.edu, wangj@njit.edu, haimin.wang@njit.edu}
\affiliation{Center for Solar-Terrestrial Research, New Jersey Institute of Technology, University Heights, Newark, NJ 07102-1982, USA}
\affiliation{Big Bear Solar Observatory, New Jersey Institute of Technology, 40386 North Shore Lane, Big Bear City, CA 92314-9672, USA}

\author{Jason T. L. Wang}
\affiliation{Institute for Space Weather Sciences, New Jersey Institute of Technology, University Heights, Newark, NJ 07102-1982, USA;
hj78@njit.edu, wangj@njit.edu, haimin.wang@njit.edu}
\affiliation{Department of Computer Science, New Jersey Institute of Technology, University Heights, Newark, NJ 07102-1982, USA}

\author{Haimin Wang}
\affiliation{Institute for Space Weather Sciences, New Jersey Institute of Technology, University Heights, Newark, NJ 07102-1982, USA;
hj78@njit.edu, wangj@njit.edu, haimin.wang@njit.edu}
\affiliation{Center for Solar-Terrestrial Research, New Jersey Institute of Technology, University Heights, Newark, NJ 07102-1982, USA}
\affiliation{Big Bear Solar Observatory, New Jersey Institute of Technology, 40386 North Shore Lane, Big Bear City, CA 92314-9672, USA}

\begin{abstract}
We present a new deep learning method, dubbed FibrilNet,
for tracing chromospheric fibrils in H$\alpha$ images of solar observations.
Our method consists of a data pre-processing component that prepares training data from a threshold-based tool, 
a deep learning model implemented as a Bayesian convolutional neural network 
for probabilistic image segmentation 
with uncertainty quantification to predict fibrils,
and a post-processing component containing a fibril-fitting algorithm
to determine fibril orientations. 
The FibrilNet tool is applied to 
high-resolution H$\alpha$ images from an active region (AR 12665) 
collected by the 1.6 m Goode Solar Telescope (GST)
equipped with high-order adaptive optics at the Big Bear Solar Observatory (BBSO). 
We quantitatively assess the FibrilNet tool, 
comparing its image segmentation algorithm and fibril-fitting algorithm 
with those employed by the threshold-based tool.
Our experimental results and major findings are summarized as follows.
First, the image segmentation results (i.e., detected fibrils) 
of the two tools are quite similar,
demonstrating the good learning capability of FibrilNet.
Second, FibrilNet finds more accurate and smoother fibril orientation angles than the threshold-based tool.
Third, FibrilNet is faster than the threshold-based tool and the uncertainty maps produced by FibrilNet 
not only provide a quantitative way to measure the confidence on each detected fibril, 
but also help identify fibril structures that are not detected by the threshold-based tool 
but are inferred through machine learning.
Finally, we apply FibrilNet to full-disk H$\alpha$ images from other solar observatories 
and additional high-resolution H$\alpha$ images collected by BBSO/GST, 
demonstrating the tool's usability in diverse datasets.
\end{abstract}


\section{Introduction} 
\label{sec:intro}

Fibrils are thin threadlike absorption features ubiquitously observed in the solar chromosphere. 
Depending on their location and dynamic behavior, they may have different names, 
e.g., threads of filaments \citep{1998SoPh..182..107M, Wang_2000}, 
the superpenumbra of sunspots \citep{1968SoPh....5..489L, Jing_2019}, 
mottles in quiet-Sun rosette structures \citep{1994A&A...282..939H}, etc. 
Fibrils are often observed with narrowband solar filtergrams in the chromospheric spectral lines such as H$\alpha$,
where they are denser than their surroundings \citep{2017A&A...607A..46M}. 
Physically speaking, fibrils represent the cool gas ``frozen" in magnetic field lines and 
protected by magnetic fields from diffusing out  \citep{1971SoPh...20..286P, 2008ApJ...679L.167L, 2009ApJ...705..272R}. 
For this reason, fibrils have been traditionally assumed to be aligned with the direction of the chromospheric magnetic field 
\citep{1971SoPh...20..298F,1971SoPh...19...59F}.

Tracing chromospheric fibrils in H$\alpha$ 
is an important subject in heliophysics research \citep{Jing_2011, Leenaarts_2015}, 
and has attracted much attention in the heliophysics community. 
The comparison between fibrils and the potential magnetic field may provide a quick way 
to examine the nonpotentiality of active regions (ARs) \citep{Jing_2011}. 
The orientation of fibrils could be used as a constraint to improve the non-linear force-free modeling of coronal fields 
\citep{2008SoPh..247..249W, 2016ApJ...826...61A, 2019ApJ...870..101F}. 
Tracing fibrils also helps estimate the amount of energy in acoustic waves 
\citep{Fossum_2006} and the free magnetic energy in the chromosphere \citep{2016ApJ...826...61A}.

Many fibril tracing methods have been developed in recent years. 
\citet{Leenaarts_2015} conducted 3-dimensional magnetohydrodynamic simulations 
to investigate the relation between chromospheric fibrils and magnetic field lines.
\citet{2016ApJ...826...61A} performed nonpotential field modeling of chromospheric structures 
and coronal loops with the VCA-NLFFF code. 
\citet{2017ApJS..229....9J} adopted the CRISPEX tool for visual inspection and identification of isolated slender fibrils. 
\citet{2017ApJS..229....6G} used image processing and contrast enhancement techniques to identify these fibrils.
\citet{2017A&A...599A.133A} employed the rolling Hough transform (RHT) for fibril detection 
and a Bayesian hierarchical model to analyze 
the pixels in spectro-polarimetric chromospheric images of penumbrae and fibrils.
The authors concluded that fibrils are often well aligned with magnetic azimuth.
This RHT technique has also been used by \citet{2017SoPh..292..132S} to analyze fibrils and coronal rain.
\citet{Jing_2011} developed a 
threshold-based algorithm 
to automatically segment chromospheric fibrils from H$\alpha$ observations
and extracted direction information along the fibrils with a fibril-fitting algorithm.
The authors further quantitatively measured the nonpotentiality of the fibrils by the magnetic shear angle. 
In contrast to the above methods, 
our deep learning-based tool (FibrilNet) presented here can automatically predict fibrils and 
measure the uncertainties in the predicted results simultaneously.

Deep learning is a branch of machine learning where neural networks are designed 
to learn from large amounts of data  \citep{LeCun2015}. 
It has been used extensively in computer vision and natural language processing, 
and more recently in astronomy and astrophysics 
for flare prediction, spectroscopic analysis, solar image segmentation, among others
\citep{Huertas_Company_2018, 10.1093/mnras/sty3217, Kim2019, 10.1093/mnras/stz761, 2019ApJ...877..121L, 10.1093/mnras/stz333, Jiang_2020}. 
Different from the previous solar image segmentation techniques,
which focus on predicting a value for each pixel, 
our FibrilNet employs a probabilistic segmentation model, specifically a Bayesian convolutional network, 
that predicts a value for each pixel accompanied with reliable uncertainty quantification.
Such a model leads to a more informed decision, and improves the quality of prediction.

In general, there are two types of uncertainty in Bayesian modeling: 
aleatoric uncertainty and epistemic uncertainty \citep{10.5555/3295222.3295309}.
Aleatoric uncertainty, also known as data uncertainty, measures the noise inherent in observations.
Epistemic uncertainty, on the other hand, measures the uncertainty in the parameters of a model; 
this uncertainty is also known as model uncertainty.
Quantifying uncertainties with machine learning finds many applications 
ranging from computer vision \citep{10.5555/3295222.3295309},
natural language processing \citep{DBLP:conf/aaai/XiaoW19}, 
medical image analysis \citep{KWON2020106816} to geomagnetic storm forecasting \citep{GCS-2018,XHY-2020}.
Here we present a new application of uncertainty quantification with machine learning in fibril tracing.

The rest of this paper is organized as follows.
Section 2 describes solar observations and data used in this study.
These data are from an active region (AR 12665) collected by the Big Bear Solar Observatory (BBSO).
Section 3 presents details of our FibrilNet method and algorithms used by the method. 
FibrilNet employs a Bayesian convolutional network for probabilistic image segmentation 
with uncertainty quantification to predict fibrils.
It then uses a fibril-fitting algorithm with a polynomial regression function of varying degrees 
to model the predicted fibrils and determine their orientations.
Section 4 reports experimental results, showing traced fibrils in the H$\alpha$ images of the solar observations 
in AR 12665 collected by BBSO.
Furthermore, we apply FibrilNet to other types of observations, 
demonstrating the tool's usability in diverse datasets.
Section 5 presents a discussion and concludes the paper.

\section{Observations and Data Preparation}
\label{sec:observational data}

The Goode Solar Telescope (GST) is a 1.6 m clear aperture, off-axis telescope at BBSO, 
which is located in Big Bear Lake, California 
\citep{2010AN....331..636C, 2010ApJ...714L..31G, 2012SPIE.8444E..03G, 2014SPIE.9147E..5DV}. 
GST is equipped with a high-order adaptive optics system, AO-308, 
which provides high-order correction of atmospheric seeing within an isoplanatic patch 
(about 6\arcsec \hspace*{+0.015cm} at 500 nm in summer), 
with a gradual roll-off of correction at larger distances \citep{2014SPIE.9148E..35S}.   
Under a stable seeing condition, BBSO/GST observed AR 12665 at (W$27^{\circ}$, S$4^{\circ}$)  
on July 13, 2017, in which the data taken during $\sim$20:16-22:42 UT were used in the study presented here.

The Visible Imaging Spectrometer \citep[VIS;][]{2010AN....331..636C} of GST
utilizes a telecentric mount of the Fabry-Pérot etalon.
This imaging system was used for observing the H$\alpha$ line. 
It scanned the target area at 
±0.6, ±0.4 and 0.0 \AA \hspace*{+0.015cm}  (0.08 \AA \hspace*{+0.015cm} bandpass) 
from the H$\alpha$ line center 6563 \AA \hspace*{+0.015cm} with a 70\arcsec \hspace*{+0.015cm} circular 
field of view (FOV). 
At each wavelength step, the 25 frames, out of 60 frames taken in succession, 
with the best contrast were saved. 
These frames, with exposure time ranging from 7 to 20 ms and an image scale of 0\arcsec.03 per pixel, 
were processed by the high-order AO system and post-facto speckle image reconstruction algorithms \citep{2008A&A...488..375W}, 
which improved the quality of the images by correcting the wavefront deformation caused by atmospheric distortion.
An H$\alpha$ line scan was performed over the FOV, and the position with the minimum intensity 
was defined as the H$\alpha$ line center. 
It should be pointed out that the GST narrowband H$\alpha$ data do not contain the full spectral information, 
which restricts the full characterization of fibrils in three dimensions.  
Therefore, our study of fibrils is all based on their projected morphology on the observational image plane.

\begin{figure}
\epsscale{1.15}
\plotone{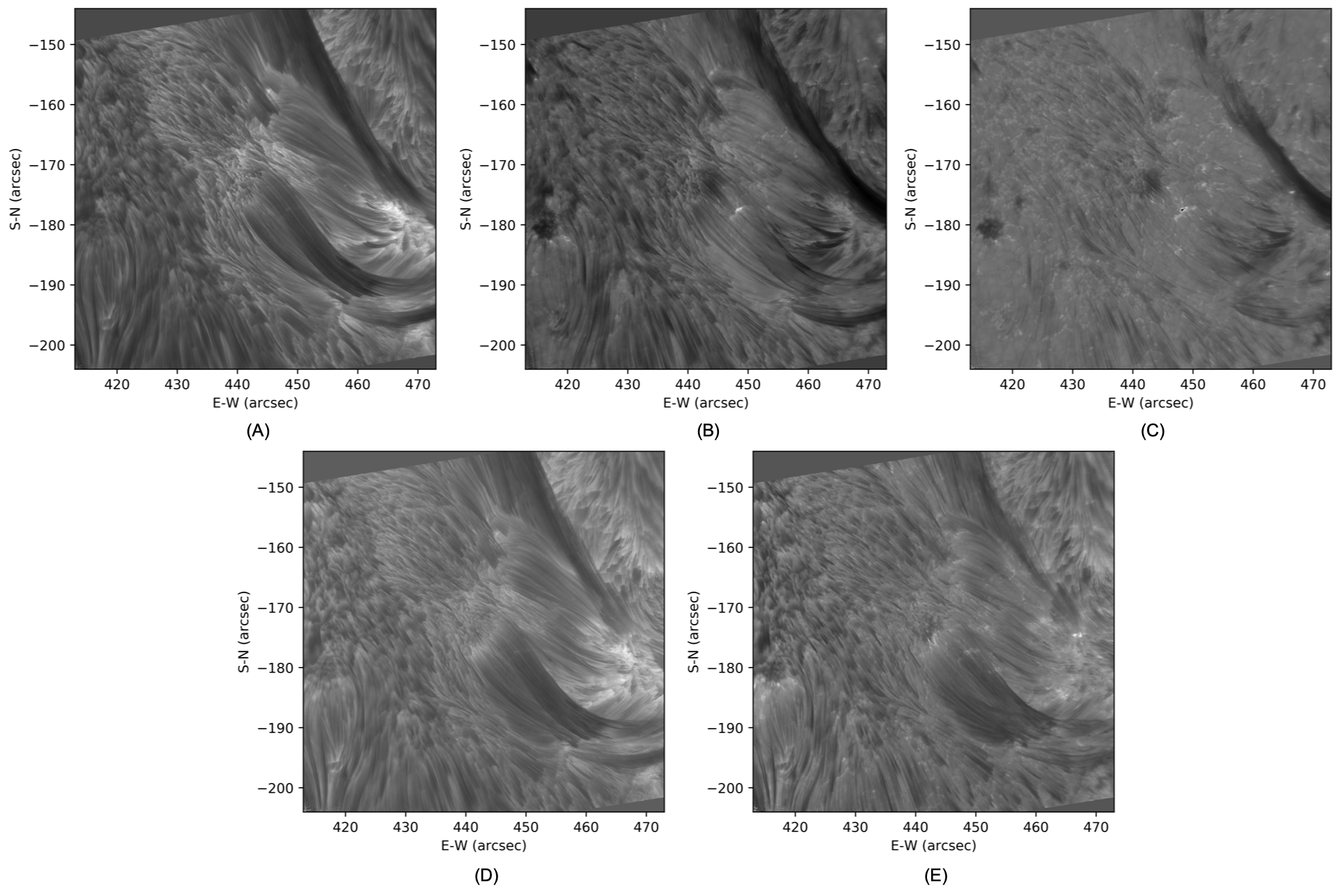}
\caption{Five test  images at 
(A) 0.0 \AA, (B) +0.4 \AA, (C) +0.6 \AA, (D) $-$0.4 \AA, (E) $-$0.6 \AA, respectively,
from the H$\alpha$ line center 6563 \AA \hspace*{+0.015cm} with a 70\arcsec \hspace*{+0.015cm}  circular FOV
collected in AR 12665 on 2017 July 13 20:15:58 UT.
Enormous amounts of fibrils exist in these H$\alpha$ images.}
\label{fig:data_overview}
\end{figure}

Our dataset contained the GST H$\alpha$ observations in AR 12665 
from 20:16:32 UT to 22:41:30 UT on July 13, 2017
where the observed region was located at (W$27^{\circ}$, S$4^{\circ}$). 
During this period of time, 241 H$\alpha$ line center images 
(i.e., those at 0.0 \AA \hspace*{+0.015cm} from the H$\alpha$ line center 6563 \AA \hspace*{+0.015cm} 
with a 70\arcsec \hspace*{+0.015cm}  circular FOV)
were used as training data since features in these images were abundant.
The test set contained five H$\alpha$ images 
taken from AR 12665 at 20:15:58 UT on the same day 
(see Figure \ref{fig:data_overview}).
Thus, there were 241 training H$\alpha$  images and 5 test H$\alpha$ images
where the size of each image was $720 \times 720$ pixels.
The training and test sets were disjoint, as the training observations and test observations were taken at different time points.
Please note that the five test images were chosen in such a way that 
they were on five different wavelength positions rather than at five different time points on the same wavelength position. 
The reason why we did not choose the test images equally distributed over the time series on the same wavelength position 
was because the features in the images on the same wavelength position did not change much across the images.
By contrast, the features in the images on the five different wavelength positions appeared quite differently 
as shown in Figure \ref{fig:data_overview}.

\section{Methodology} 
\label{sec:method}

\subsection{Overview of FibrilNet}
\label{sec:overview}

\begin{figure}
\epsscale{1.15}
\plotone{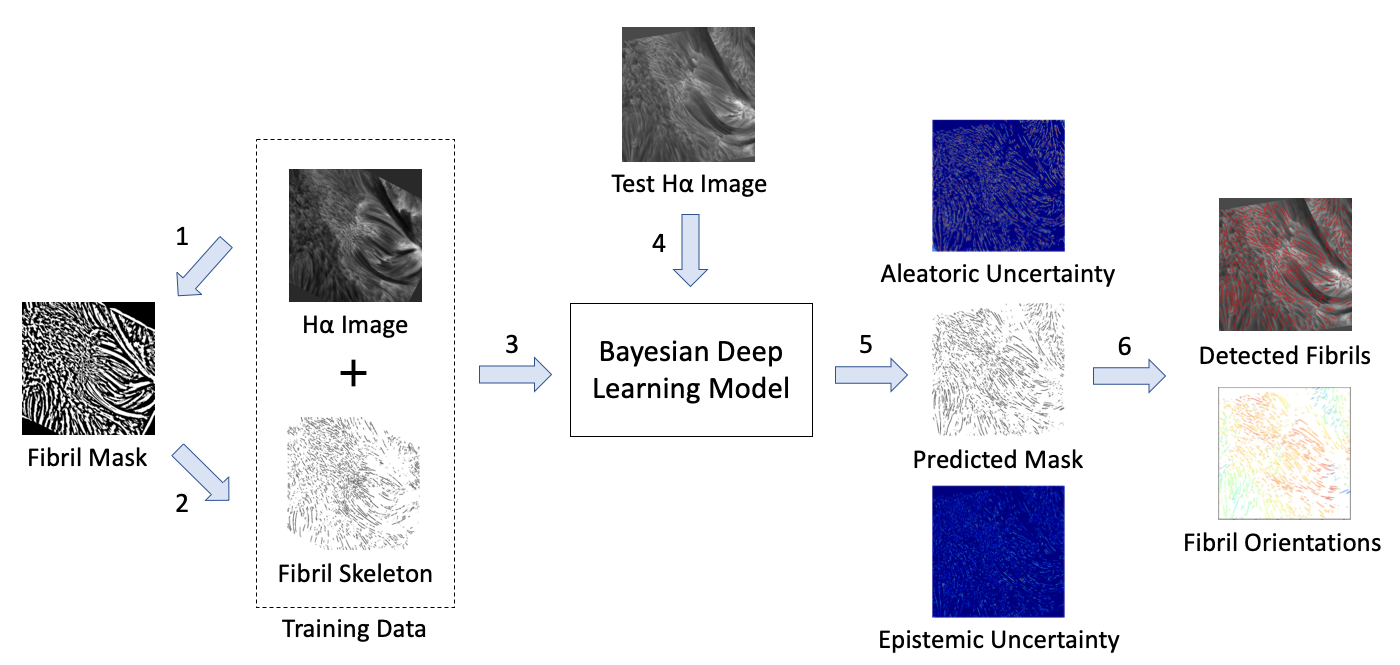}
\caption{Illustration of the proposed method (FibrilNet) for fibril tracing.
FibrilNet employs a Bayesian deep learning model for probabilistic image segmentation 
with uncertainty quantification to predict fibrils
and a fibril-fitting algorithm to determine fibril orientations.
The training data used to train the Bayesian deep learning model are highlighted in the dashed box.
The tracing results for the test H$\alpha$ image include predicted/detected fibrils, 
their orientations, aleatoric uncertainty and epistemic uncertainty.}
\label{fig:method}
\end{figure}

Figure \ref{fig:method} explains how FibrilNet works.
Training H$\alpha$ images are pre-processed in steps 1 and 2, 
and then used to train the Bayesian deep learning model (step 3).
The trained model takes as input a test H$\alpha$ image (step 4) and produces as output a predicted mask 
accompanied with results for quantifying aleatoric uncertainty and epistemic uncertainty (step 5).
In the post-processing phase (step 6), based on the predicted mask,
fibrils on the test H$\alpha$ image are detected and highlighted by thin red curves.
Furthermore, the orientations of the detected fibrils are determined based on a 
fibril-fitting algorithm where the orientations are shown by different colors.

Specifically, in step 1, we apply the threshold-based tool developed by 
\citet{Jing_2011} to each training H$\alpha$ image described in Section \ref{sec:observational data} 
to obtain a corresponding fibril mask. 
Fibril patterns on this mask are very thick, which contains a lot of noise.
In step 2, we refine the fibril mask via a skeletonization procedure
to obtain a fibril skeleton in which fibrils are marked by black 
and regions without fibrils are marked by white.
The skeletonization procedure works by extracting a region-based shape feature representing the general form of fibrils.
This skeletonization procedure results in better and cleaner images suitable for model training \citep{Umbaugh:2010:DIP:1951634}.

The training H$\alpha$ images and fibril skeletons
are then used to train the Bayesian deep learning model for probabilistic image segmentation and uncertainty quantification (step 3).
During training, in order to obtain a robust model, we use the data augmentation technique described in \citet{Jiang_2020}
to expand the training set by shifting, rotating, flipping and scaling the training images.
In step 4, a test H$\alpha$ image is 
fed to the trained Bayesian deep learning model. 
During testing, we use the Monte Carlo (MC) dropout sampling technique described in Section \ref{sec:implementation}
to produce the predicted mask of the test H$\alpha$ image accompanied with 
aleatoric uncertainty and epistemic uncertainty results (step 5).
In step 6, by using the fibril-fitting algorithm based on the polynomial regression model described in Section \ref{sec:fibril_tracing},
our FibrilNet tool outputs detected fibrils marked by red color on the test H$\alpha$ image
and their orientations represented by different colors. 

\subsection{Implementation of the Bayesian Deep Learning Model in FibrilNet}
\label{sec:implementation}

\begin{figure}
\epsscale{1.15}
\plotone{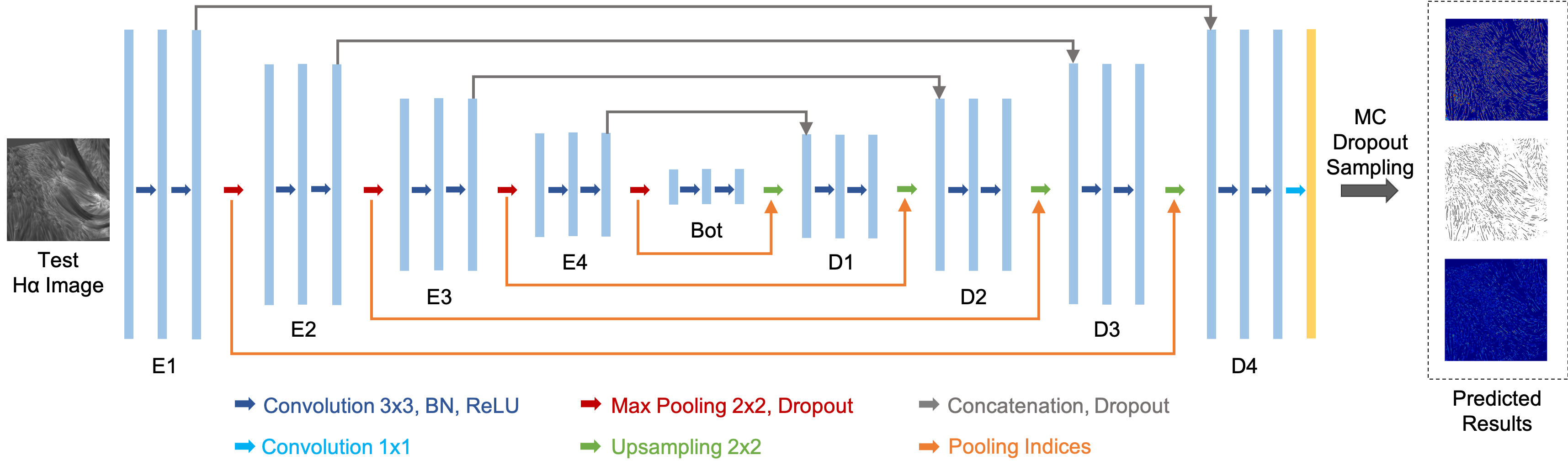}
\caption{Architecture of the encoder-decoder convolutional neural network (i.e., the Bayesian deep learning model) used in FibrilNet. 
This network is similar to the one presented in \citet{Jiang_2020}.
See text for their differences.}
\label{fig:fibrilNet}
\end{figure}

The Bayesian deep learning model used by FibrilNet
is similar to the model used in SolarUnet \citep{Jiang_2020} for tracking magnetic flux elements.
Both models have 4 encoder blocks (E1, E2, E3, E4),
4 decoder blocks (D1, D2, D3, D4), mediated by
a bottleneck (Bot).
See Figure \ref{fig:fibrilNet} and \citet{Jiang_2020}
for the configuration and parameter settings of the models.
While both models are based on an encoder-decoder convolutional neural network,
they differ in three ways.
First, in performing 2$\times$2 max pooling, represented by a red arrow in Figure \ref{fig:fibrilNet},
the corresponding max pooling indices are stored.
During decoding, the max pooling indices at the corresponding encoder layer are recalled,
represented by an orange arrow,
to upsample, represented by a green arrow, as done in \citet{7803544}.
This upsampling technique used by FibrilNet, designed to 
reduce the number of trainable parameters in the model (network) and hence save memory,
is different from the up-convolution layers used in SolarUnet.  
Second, since fibril patterns are relatively vague and harder to identify than magnetic flux elements, 
FibrilNet uses twice as many kernels as those in SolarUnet
in all of the blocks in the encoder and decoder, as well as the bottleneck.
Finally, during testing, instead of using the trained model (network) directly 
to produce segmentation results as done in SolarUnet,
FibrilNet employs a Monte Carlo (MC) dropout sampling technique, detailed below, to produce,
for a test H$\alpha$ image, a predicted mask accompanied with 
aleatoric uncertainty and epistemic uncertainty results (see Figure \ref{fig:method}).
This MC dropout sampling technique allows FibrilNet to perform probabilistic image segmentation
with uncertainty quantification, which is lacking in SolarUnet.

Specifically, to quantify uncertainty with the convolutional neural network, 
we use a prior probability, $P(\mathbf{W})$,  over the network's weights, $\mathbf{W}$.
During training, pairs of H$\alpha$ images and their corresponding fibril skeletons, 
collectively referred to as $\mathbf{D}$,
are used to train the network.
According to Bayes' theorem,
\begin{equation}
P(\mathbf{W} \mid \mathbf{D})=\frac{P(\mathbf{D} \mid \mathbf{W}) P(\mathbf{W}) }{P(\mathbf{D})}.  
\end{equation}
Computing the exact posterior probability, $P(\mathbf{W} \mid \mathbf{D})$, is intractable \citep{10.5555/2986766.2986882}.
Nevertheless, we can use variational inference \citep{10.5555/2986459.2986721} 
to learn the variational distribution over the network's weights
parameterized by $\theta$, $q_{\theta}(\textbf{W})$,
by minimizing the Kullback–Leibler (KL) divergence of $q_{\theta}(\textbf{W})$ and 
$P(\mathbf{W} \mid \mathbf{D})$ \citep{doi:10.1080/01621459.2017.1285773}.
It is known that training a network with dropout is equivalent to a variational approximation on the network 
\citep{10.5555/3045390.3045502}.
Furthermore, minimizing the cross-entropy loss of the network is equivalent to the minimization of the KL divergence  \citep{Goodfellow-et-al-2016}.
Therefore, we use a binary cross-entropy loss function and the adaptive moment estimation (Adam) optimizer \citep{Goodfellow-et-al-2016}
with a learning rate of 0.0001 to train our model (network).
Let $\hat{\theta}$ denote the optimized variational parameter obtained by training the model (network);
we use $q_{\hat{\theta}}(\mathbf{W})$ to represent the optimized weight distribution.
 
In deep learning, dropout is mainly used to prevent over-fitting,
where a trained model overfits training data and hence can not be generalized to make predictions on unseen test data.
During training, dropout refers to ignoring or dropping out units (i.e., neurons) of certain set of neurons which is chosen randomly.
During testing, dropout can be used to retrieve $T$ Monte Carlo (MC) samples by processing the input test H$\alpha$ image $T$ times 
\citep{10.5555/3045390.3045502}.
(In the study presented here, $T$ is set to 50.)
Each time a set of weights is randomly drawn from $q_{\hat{\theta}}(\mathbf{W})$.
Each pixel in the predicted mask, shown in step 5 of  Figure \ref{fig:method}, gets a mean and variance over the $T$ samples.
If the mean is greater than or equal to a threshold, the pixel is marked by black 
indicating that the pixel is part of a fibril;
otherwise the pixel is marked by white
indicating that the pixel is not part of a fibril.
(In the study presented here, the threshold is set to 0.5.)
Following \citet{KWON2020106816}, we decompose the variance into the aleatoric uncertainty and epistemic uncertainty at the pixel.
The aleatoric uncertainty captures the inherent randomness of the predicted result, which comes from the input test H$\alpha$ image, 
while the epistemic uncertainty comes from the variability of $\mathbf{W}$, 
which accounts for the uncertainty in the model parameters (weights).

In the post-processing phase,
we use a connected-component labeling algorithm \citep{He:2009:FCL:1542560.1542851} 
to group all adjacent black segments if their pixels in edges or corners touch each other.
For each resulting group, which represents a fibril,
we locate its pixels in the predicted mask and highlight
their corresponding pixels in the test H$\alpha$ image by red.
(Resulting groups containing less than 10 pixels are considered as noise and filtered out.)
We then output the detected fibrils highlighted by red color in the test H$\alpha$ image,
as shown in step 6 of Figure \ref{fig:method}.

\subsection{Implementation of the Fibril-Fitting Algorithm in FibrilNet}
\label{sec:fibril_tracing}

Most of the detected fibrils are lines or curves.
In contrast to \citet{Jing_2011}, which used a quadratic function to fit the detected fibrils,
we adopt a polynomial regression model here.
Specifically, our regression model is a polynomial function with varying degrees 
capable of fitting the detected fibrils with different curvatures.
In general, regression analysis investigates the relationship between a dependent variable and an independent variable \citep{10.5555/1162264}.
We model a detected fibril as an $nth$ degree polynomial function as follows:
\begin{equation}
y =\gamma_{0}+\gamma_{1} x+\gamma_{2} x^{2}+ \ldots + \gamma_{n} x^{n}+ \epsilon,
\label{polynomial}
\end{equation}
where $\gamma_{i}$ are coefficients and $\epsilon$ is a random error term. 
In Equation (\ref{polynomial}), 
the independent variable $x$ represents the $x$ coordinate of a pixel in the detected fibril 
and the dependent variable $y$ represents the $y$ coordinate of the same pixel, 
where the $x$-axis represents the E-W direction and the $y$-axis represents the S-N direction (see Figure \ref{fig:data_overview}).
When the degree $n$ equals 1, Equation (\ref{polynomial}) represents a linear regression model, 
meaning that the detected fibril is represented by a straight line.
In our work, $n$ ranges from 1 to 10.

We then use the least squares method \citep{OSTERTAGOVA2012500} to find the optimal $\gamma_{i}$ values.
There are 10 candidate polynomial functions for representing the detected fibril.
We use the R-squared score \citep{OSTERTAGOVA2012500} to assess the feasibility of these 10 candidate polynomial functions.
Specifically, we choose the candidate polynomial function yielding the largest R-squared score, and
use this polynomial function to represent the detected fibril.
To determine the orientation of the detected fibril, we calculate the derivative of the chosen polynomial function.
For each pixel on the detected fibril, we thus obtain the slope of the tangent at the pixel, leading to the
orientation angle of the pixel, denoted $\theta_{f}$, with respect to the $x$-axis.
Notice that the orientation angle $\theta_{f}$ is in the $0^{\circ}-180^{\circ}$ range,
as two directions differing by $180^{\circ}$ are indistinguishable here because the detected fibril in H$\alpha$ does not carry
information on the vertical dimension.
Thus, $\theta_{f}$ represents the direction of the detected fibril with a $180^{\circ}$ ambiguity  \citep{Jing_2011}. 

\section{Results} 
\label{sec:experiment}

\subsection{Tracing Results of FibrilNet Based on Data from AR 12665}
\label{Tracing Halpha results}

\begin{figure}
\epsscale{1.15}
\plotone{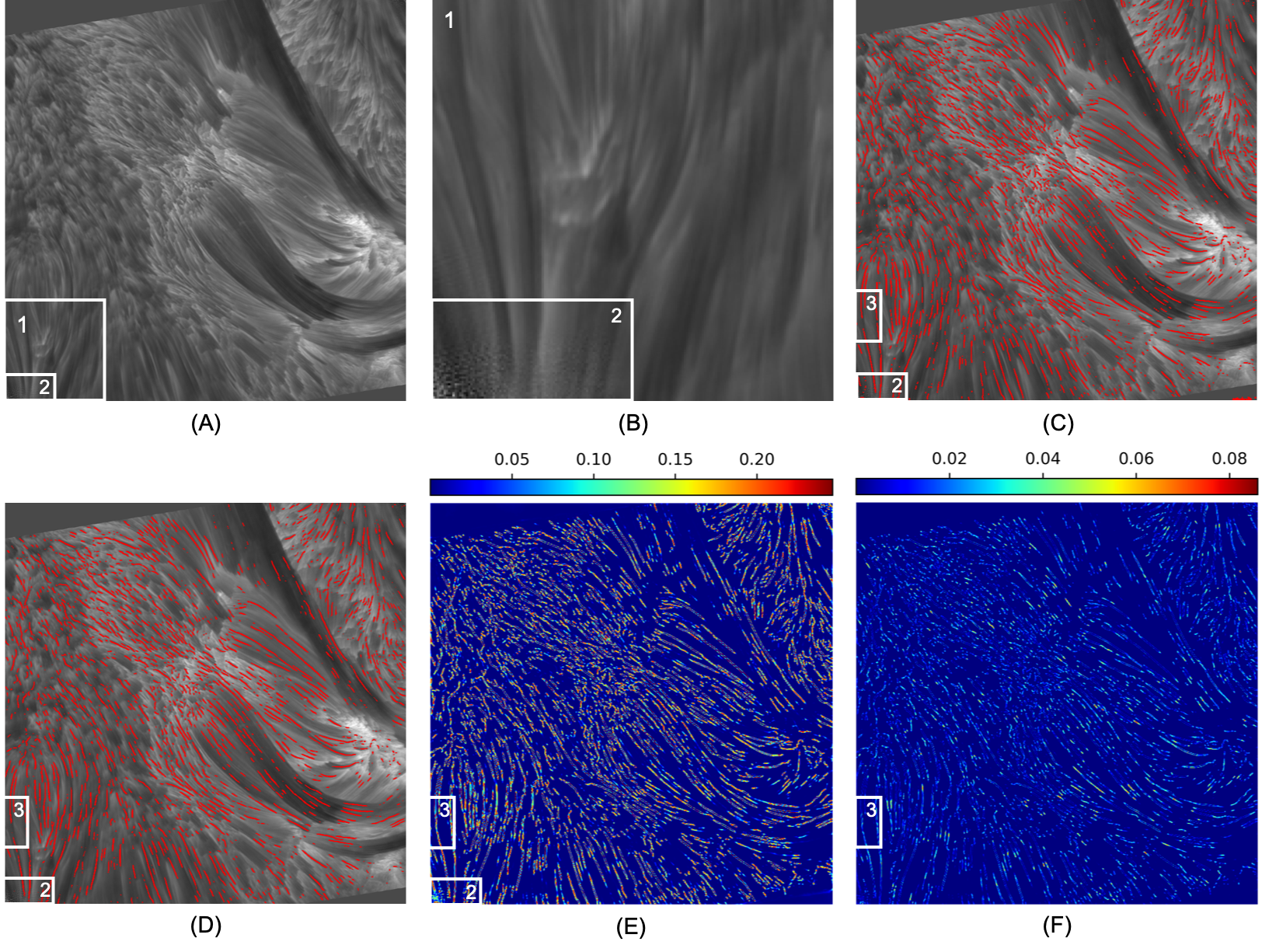}
\caption{Fibril tracing results on the test image 
at 0.0 \AA \hspace*{+0.015cm} from the H$\alpha$ line center 6563 \AA \hspace*{+0.015cm} 
with a 70\arcsec \hspace*{+0.015cm}  circular FOV 
collected in AR 12665 on 2017 July 13 20:15:58 UT
where training data were 241 H$\alpha$ line center images taken from the same AR 
between 20:16:32 UT and 22:41:30 UT on the same day.
(A) The original test H$\alpha$ image.
(B) The enlarged FOV of the region highlighted by the white box 1 in (A).
(C) Fibrils (red curves) on the test H$\alpha$ image detected by the tool in \citet{Jing_2011}.
(D) Fibrils (red curves) on the test H$\alpha$ image predicted by FibrilNet.
(E) The aleatoric uncertainty (data uncertainty) map produced by FibrilNet.
(F) The epistemic uncertainty (model uncertainty) map produced by FibrilNet. 
\label{fig: model_results}}
\end{figure}

In this series of experiments, we used the 241 H$\alpha$ line center images from 20:16:32 UT to 22:41:30 UT on 2017 July 13 
mentioned in Section \ref{sec:observational data} 
along with their corresponding fibril skeletons to train the FibrilNet tool as described in Section \ref{sec:method}.
We then used the trained tool to predict and trace fibrils on the five test  images at
0.0 \AA, +0.4 \AA, +0.6 \AA, $-$0.4 \AA, $-$0.6 \AA, respectively,
from the H$\alpha$ line center 6563 \AA \hspace*{+0.015cm} with a 70\arcsec \hspace*{+0.015cm}  circular FOV 
collected in AR 12665 on 2017 July 13 20:15:58 UT
(see Figure \ref{fig:data_overview}).
Figure \ref{fig: model_results} presents tracing results on the test image at 0.0 \AA;
tracing results on the other four test images can be found in the Appendix.
In all of the tracing results, fibrils containing 10 or fewer pixels were treated as noise and excluded.

Figure \ref{fig: model_results}(A) shows the original test H$\alpha$ image.
Figure \ref{fig: model_results}(B) shows the enlarged FOV of the region highlighted by 
the white box 1
in Figure \ref{fig: model_results}(A).
It can be seen from Figure \ref{fig: model_results}(B) that there are salt-and-pepper noise pixels in the region 
highlighted by the white box 2, where the noise pixels are caused by image reconstruction limitations.
Figure \ref{fig: model_results}(C) shows the fibrils (red curves) on the test H$\alpha$ image
detected by the tool (after skeletonization) presented in \citet{Jing_2011}.
Figure \ref{fig: model_results}(D) shows the fibrils (red curves) predicted by FibrilNet.
FibrilNet uses the images processed by the tool in \citet{Jing_2011} as training data.
The results in Figures \ref{fig: model_results}(C) and \ref{fig: model_results}(D) are quite similar,
demonstrating the good learning capability of FibrilNet.  

Figures \ref{fig: model_results}(E) and \ref{fig: model_results}(F) show the 
aleatoric uncertainty (data uncertainty) and epistemic uncertainty (model uncertainty) maps, respectively, produced by FibrilNet.
Regions predicted with less uncertainty and higher confidence are colored by blue.
Regions predicted with more uncertainty and lower confidence are colored by red.
We can see that the main source of uncertainty comes from the data rather than the model.
Specifically, the values in the data uncertainty map in Figure \ref{fig: model_results}(E) range from 0 to 0.246 
while the values in the model uncertainty map in Figure \ref{fig: model_results}(F) range from 0 to 0.086.
Furthermore, we observe that the ends of a detected fibril are often associated with higher uncertainty.
This happens because there is ambiguity surrounding the transition from the fibril body to the non-fibril background area,
a finding consistent with that in object detection with uncertainty quantification reported in the literature 
\citep{10.5555/3295222.3295309,KWON2020106816}.

Notice also that the map in Figure \ref{fig: model_results}(E) shows higher uncertainty in the noisy region inside the white box 2
compared to the region outside the white box 2.
Specifically, in the noisy region inside the white box 2, 
90\% of the values are contained in the range [0.00014 (5\%), 0.20528 (95\%)]. 
By contrast, in the region outside the white box 2, 90\% of the values are contained in the range [0 (5\%), 0.18969 (95\%)].
Furthermore, it is observed in 
Figure \ref{fig: model_results}(C) 
that the tool in \citet{Jing_2011} misses some fibril structures with at least 15 pixels
inside the white box 3. 
These fibril structures are not present in the mask predicted by FibrilNet either,
as shown inside the white box 3  in Figure \ref{fig: model_results}(D).
Nevertheless,  the uncertainty maps of FibrilNet are able to catch and display 
these missed fibril structures with higher uncertainty by Bayesian inference,
as shown inside the white box 3 in Figures \ref{fig: model_results}(E) and \ref{fig: model_results}(F) respectively.
This finding demonstrates the usefulness of the uncertainty maps, as they 
not only provide a quantitative way to measure the confidence on each predicted fibril, 
but also help identify fibril structures that are not detected by the tool in \citet{Jing_2011}  
but are inferred through machine learning.
It should be pointed out that the previous fibril tracing tool in \citet{Jing_2011}
does not have the capability of producing these uncertainty maps as described here. 

\begin{figure}
\epsscale{1.18}
\plotone{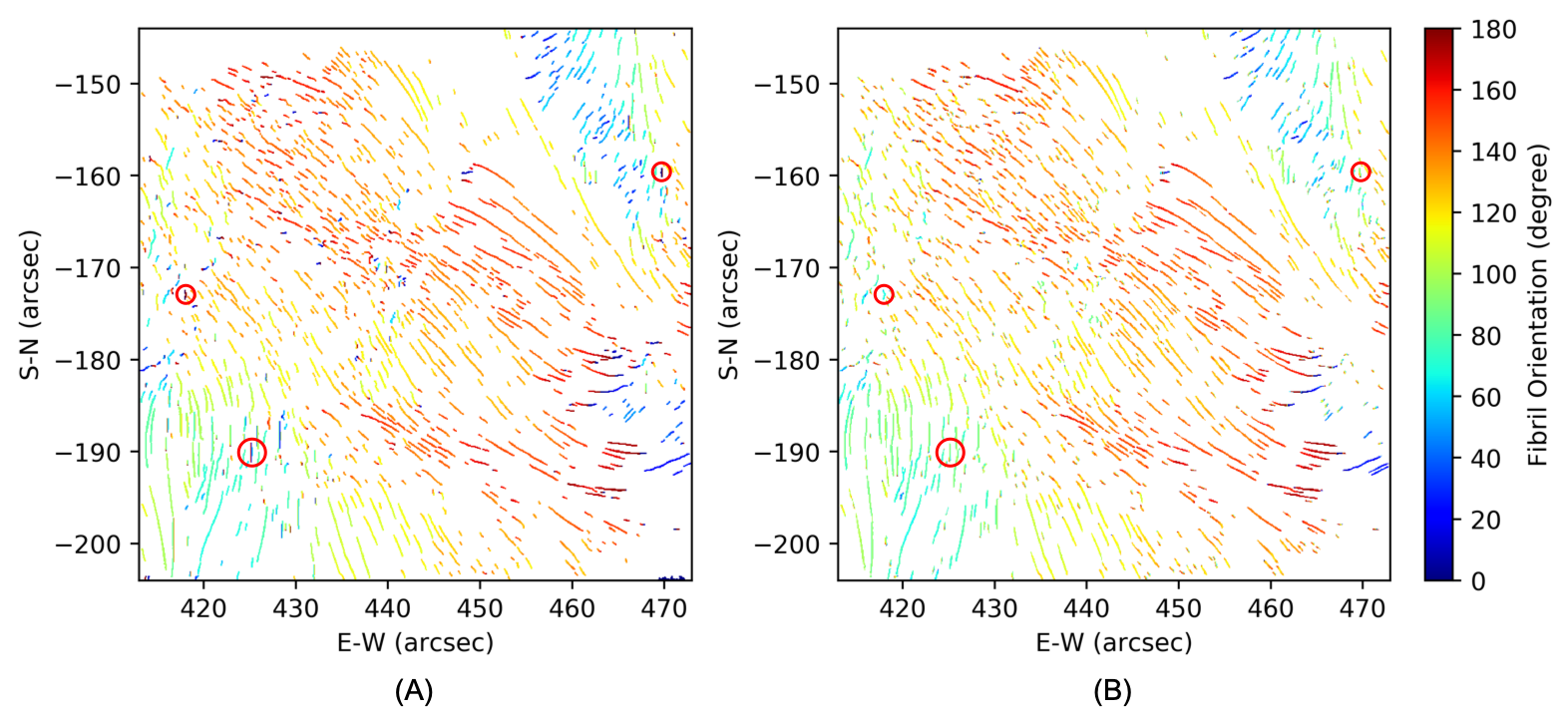}
\caption{Orientation angles (colored curves) of the detected fibrils on the test image 
at 0.0 \AA \hspace*{+0.015cm} from the H$\alpha$ line center 6563 \AA \hspace*{+0.015cm} 
with a 70\arcsec \hspace*{+0.015cm}  circular FOV 
collected in AR 12665 on 2017 July 13 20:15:58 UT.
(A) Fibril orientation angles calculated by the tool in \citet{Jing_2011}.
(B) Fibril orientation angles determined by FibrilNet.
Orientation angles of a number of fibrils,
some of which are highlighted by small red circles here,
are calculated wrongly by the tool in \citet{Jing_2011}, but correctly by FibrilNet.
\label{fig: orientation}}
\end{figure}

Figure \ref{fig: orientation} compares the orientation angles of the fibrils found by the tool in \citet{Jing_2011} 
and by FibrilNet respectively.
The colors of angles between $0^{\circ}$ and $90^{\circ}$ range from dark blue to green.
The colors of angles between $90^{\circ}$ and $180^{\circ}$ range from green to dark red.
It can be seen from Figure \ref{fig: orientation} that 
the orientation angles found by the two tools mostly agree with each other, 
though the angles detected by FibrilNet tend to be smoother.
This happens because FibrilNet uses polynomial functions of varying degrees,
as opposed to the quadratic function employed by the tool in \citet{Jing_2011},
to better fit the detected fibrils with different curvatures.
Notice also that the quadratic function used by the tool in \citet{Jing_2011}
may produce wrong angles, which are calculated correctly by the polynomial regression model of FibrilNet.
For example, the orientation angle of the fibril at
E-W = 425\arcsec\ and S-N = $-190$\arcsec,
which is highlighted by a small red circle
in Figure \ref{fig: orientation}, is roughly $90^{\circ}$.
It is calculated incorrectly by the tool in \citet{Jing_2011}, 
as shown in Figure \ref{fig: orientation}(A).
On the other hand, FibrilNet calculates the orientation angle of this fibril correctly,
as shown in Figure \ref{fig: orientation}(B).
It should be pointed out that the smoother and more accurate orientation angles
detected by FibrilNet are due to the better fibril-fitting algorithm used by the tool, as explained above.
They are not caused by FibrilNet's Bayesian deep learning model, whose purpose is mainly for image segmentation
(i.e., marking each pixel by black indicating the pixel is part of a fibril or white indicating the pixel is not part of a fibril 
as shown in the predicted mask in Figure \ref{fig:method})
with uncertainty quantification (i.e., producing the uncertainty maps as shown in Figure \ref{fig: model_results}).

\subsection{Quantitative Assessment of FibrilNet Based on Data from AR 12665}
\label{sec:quantitative}

As mentioned above, FibrilNet has two parts: 
(i) the Bayesian deep learning model for predicting fibrils with probabilistic image segmentation, and
(ii) the fibril-fitting algorithm for determining orientations of the predicted fibrils 
based on the polynomial regression function in Equation (\ref{polynomial}). 
Here, we adopt four measures, defined below, to quantitatively assess the first part,
comparing the image segmentation algorithms employed by FibrilNet and the tool (after skeletonization) in \citet{Jing_2011}, 
based on the same data from AR 12665 used in Section \ref{Tracing Halpha results}.
Unlike FibrilNet, which employs deep learning for image segmentation,
the tool in \citet{Jing_2011} used a 
threshold-based algorithm
rather than machine learning for image segmentation.

Let $A$ ($B$, respectively) denote the set of $720 \times 720 = 518,400$ pixels 
in the mask (skeleton, respectively)
predicted by FibrilNet (calculated by the tool in \citet{Jing_2011}, respectively) for a test image.
Let $p \in A$ be a pixel in $A$ and let $q \in B$ be $p$'s corresponding pixel in $B$,
i.e., $q$ is at the same position as $p$.
We use $A \cap_{A} B$ to represent the subset of pixels in $A$ 
such that for each pixel $p$ in $A \cap_{A} B$ and $p$'s corresponding pixel $q$ in $B$,
$p$ and $q$ are marked by the same color.
That is, $p$, $q$ are both marked by black indicating $p$ ($q$, respectively) is part of a fibril in $A$ ($B$, respectively),
or $p$, $q$ are both marked by white indicating $p$ ($q$, respectively) is not part of a fibril in $A$ ($B$, respectively).
Similarly, we use $A \cap_{B} B$ to represent the subset of pixels in $B$
such that for each pixel $q$ in $A \cap_{B} B$ and $q$'s corresponding pixel $p$ in $A$,
$q$ and $p$ are marked by the same color.
The first quantitative measure is the pixel similarity (PS), also known as global accuracy \citep{7803544}, 
which is defined as
\begin{equation}
\operatorname{PS} = \frac{|A \cap_{A} B| +  |A \cap_{B} B|}{|A| + |B|},
\end{equation}
where $|.|$ is the cardinality of the indicated set.
PS is used to assess the pixel-level similarity between the mask $A$ predicted by FibrilNet and
the skeleton $B$ calculated by the tool in \citet{Jing_2011} for the test image.
The value of PS ranges from 0 to 1.
The larger (i.e., closer to 1) the PS value, the higher the pixel-level similarity between the mask $A$ 
and the skeleton $B$.

Let $A_{F}$ ($B_{F}$, respectively) denote the set of pixels on the fibrils in $A$ ($B$, respectively).
Thus, in $A$, the pixels in $A_{F}$ are marked by black while the pixels not in $A_{F}$ are marked by white.
Similarly, in $B$, the pixels in $B_{F}$ are marked by black while the pixels not in $B_{F}$ are marked by white.
We use $A_{F} \cap_{A_{F}} B_{F}$ to represent the subset of pixels in $A_{F}$ 
such that for each black pixel $p$ in $A_{F} \cap_{A_{F}} B_{F}$, $p$'s corresponding pixel $q$ is also black, i.e., $q$ is in $B_{F}$.
Similarly, we use $A_{F} \cap_{B_{F}} B_{F}$ to represent the subset of pixels in $B_{F}$ 
such that for each black pixel $q$ in $A_{F} \cap_{B_{F}} B_{F}$, $q$'s corresponding pixel $p$ is also black, i.e., $p$ is in $A_{F}$.
The second quantitative measure is the fraction of common fibril pixels (FCFP), defined as
\begin{equation}
\operatorname{FCFP} = \frac{|A_{F} \cap_{A_{F}} B_{F}| +  |A_{F} \cap_{B_{F}} B_{F}|}{|A_{F}| + |B_{F}|}.
\end{equation}
FCFP is used to measure the pixel-level similarity between the fibrils predicted by FibrilNet 
and those found by the tool in \citet{Jing_2011}.
The value of FCFP ranges from 0 to 1.
The larger (i.e., closer to 1) the FCFP value, the higher the pixel-level similarity 
between the fibrils predicted by FibrilNet and those found by the tool in \citet{Jing_2011}.

The third quantitative measure is the fraction of disjunct fibril pixels (FDFP), defined as
\begin{equation}
\operatorname{FDFP} = 1 - \operatorname{FCFP}.
\end{equation}
FDFP is used to measure the pixel-level dissimilarity (distance) between the fibrils predicted by FibrilNet 
and those found by the tool in \citet{Jing_2011}.
The value of FDFP ranges from 0 to 1.
The smaller (i.e., closer to 0) the FDFP value, the higher the pixel-level similarity 
between the fibrils predicted by FibrilNet and those found by the tool in \citet{Jing_2011}.

The fourth quantitative measure is the Rand Index \citep[RI;][]{10.2307/2284239, 1565332},
which calculates the ratio of pairs of pixels whose 
colors (black or white) 
are consistent between the mask $A$ predicted by FibrilNet and
the skeleton $B$ calculated by the tool in \citet{Jing_2011} for the test image.
RI accommodates the inherent ambiguity in image segmentation, 
and provides region sensitivity and compensation for coloring errors near
the ends of detected fibrils.
For example, consider a wider fibril.
FibrilNet may detect the portion to the left of the center of the fibril and highlight this portion by red.
The tool in \citet{Jing_2011} may detect the portion to the right of the center of the fibril and highlight that portion by red.
Under this circumstance, FCFP does not consider there are common pixels between the two red curves,
though RI treats the two red curves as consistent curves.
Visually the fibril is indeed found by both tools.
As a consequence, RI is often used in comparing image segmentation algorithms.
The value of RI also ranges from 0 to 1.
The larger (i.e., closer to 1) the RI value, the higher the visual similarity 
between the fibrils predicted by FibrilNet and those found by the tool in \citet{Jing_2011}.

Table \ref{table: metrics} presents the quantitative measure values of FibrilNet 
based on the five test images in Figure \ref{fig:data_overview}.
It can be seen from the table that the mask predicted by FibrilNet and the skeleton calculated by the tool in \citet{Jing_2011}
are very similar at pixel level, with PS $\geq 95\%$ on the test images.
The fraction of common fibril pixels (FCFP) is about $80\%$.
However, visually, the similarity/consistency between the fibrils predicted by FibrilNet 
and those found by the tool in \citet{Jing_2011} is much higher,
where the similarity/consistency is quantitatively assessed with RI $\geq 91\%$ on the test images.
This finding is consistent with the results presented in Figures \ref{fig: model_results}(C) and \ref{fig: model_results}(D).

\begin{table}
\centering
\caption{Comparison of the Image Segmentation Algorithms Used in FibrilNet and the Tool of \citet{Jing_2011} 
Based on Four Quantitative Measures and Five Test Images}
\begin{tabular*}{0.7\textwidth}{@{\extracolsep{\fill}} l c c c c  }
\hline
\hline
Test Image & PS & FCFP & FDFP & RI \\
\hline
H$\alpha$ 0.0 \AA  & 0.9576 & 0.8038 & 0.1962 &  0.9188   \\	
H$\alpha$ +0.4 \AA& 0.9571 & 0.8097 & 0.1903 &  0.9178  \\
H$\alpha$ +0.6 \AA& 0.9659 & 0.8079 & 0.1921 &  0.9340  \\
H$\alpha$ $-$0.4 \AA & 0.9546 & 0.7922 & 0.2078 &  0.9134  \\
H$\alpha$ $-$0.6 \AA & 0.9536 & 0.8022 & 0.1978 &  0.9115  \\
\hline	
\end{tabular*}
\label{table: metrics}
\end{table}
	
Next, we quantitatively assess the second part of FibrilNet,
comparing the fibril-fitting algorithms employed by FibrilNet and the tool (after skeletonization) in \citet{Jing_2011}, 
based on the same data from AR 12665 described in Section \ref{Tracing Halpha results}.
The fibril-fitting algorithms are used to determine orientations of detected fibrils.
Let $\theta_{f}$ represent the fibril orientation angle of a pixel 
calculated by the polynomial regression function in FibrilNet, and 
let $\theta_{j}$ represent the fibril orientation angle of
the same pixel calculated by the quadratic function in the tool of \citet{Jing_2011}.
The acute angle difference between $\theta_{f}$ and $\theta_{j}$, denoted $\delta$($\theta_{f}$, $\theta_{j}$), 
is defined as
\begin{equation}
\delta(\theta_{f}, \theta_{j}) = \left\{ \begin{array}{ll}
|\theta_{f} - \theta_{j}|                        & \mbox{if $|\theta_{f} - \theta_{j}|$ $\leq$ $90^{\circ}$}   \\
180^{\circ} - |\theta_{f} - \theta_{j}|  &  \mbox{otherwise} 
\end{array}
\right..
\label{angle_dif}
\end{equation}
The angle difference 
is decided in favor of an acute or right angle, i.e., 
$0^{\circ}$ $\leq$ $\delta(\theta_{f}, \theta_{j})$ $\leq$ $90^{\circ}$. 

\begin{figure}
\epsscale{1.15}
\plotone{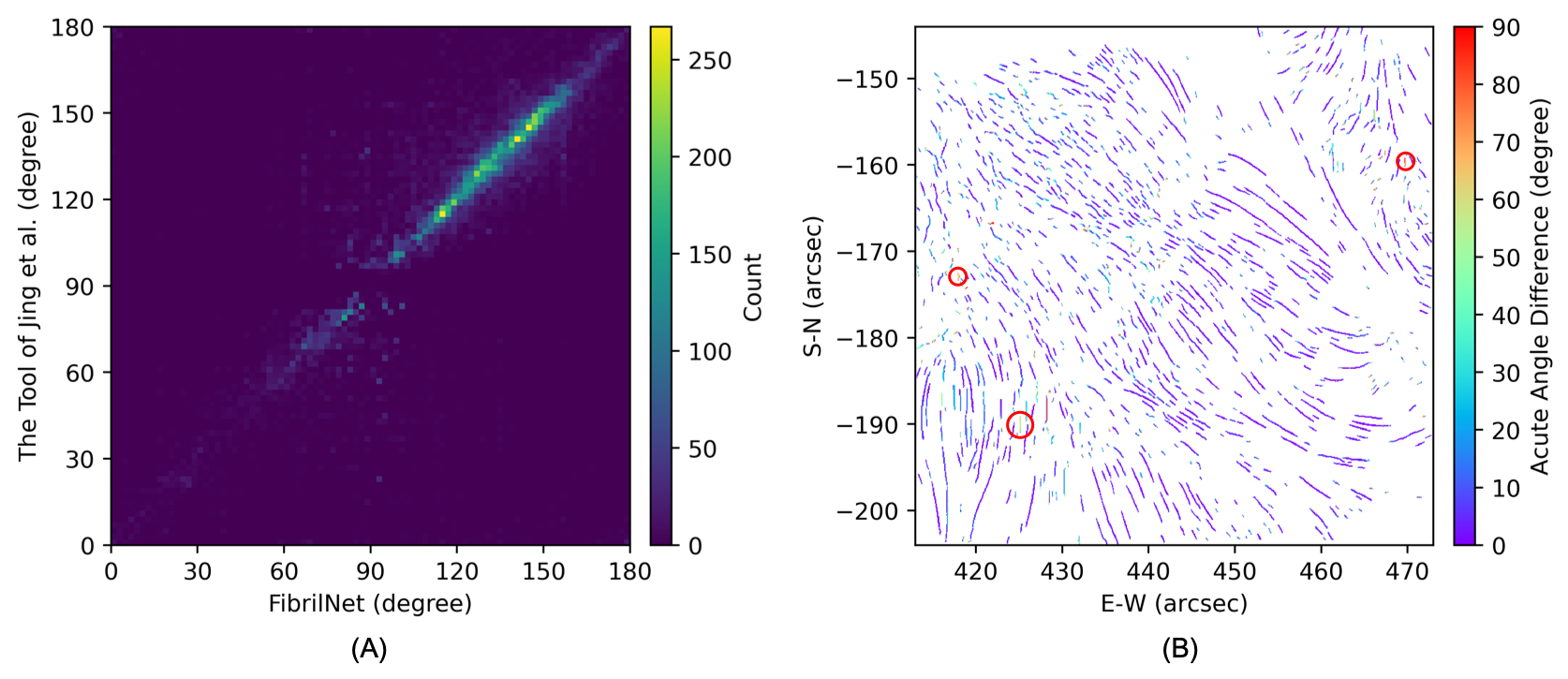}
\caption{Quantitative comparison of the orientation angles of common fibril pixels calculated by the fibril-fitting algorithms 
used in FibrilNet and the tool of \citet{Jing_2011}
based on the test image at 0.0 \AA \hspace*{+0.015cm} from the H$\alpha$ line center 6563 \AA \hspace*{+0.015cm} 
with a 70\arcsec \hspace*{+0.015cm}  circular FOV collected in AR 12665 on 2017 July 13 20:15:58 UT.
(A) 2D histogram of the orientation angles of common fibril pixels produced by the two tools.
(B) Differences of the orientation angles of common fibril pixels produced by the two tools.
The orientation angles of 
the fibrils highlighted by small red circles
are calculated wrongly by the tool of \citet{Jing_2011}, 
but correctly by FibrilNet as indicated in Figure \ref{fig: orientation}.
\label{fig: common_pixels_angle_diff} }
\end{figure}

Figure \ref{fig: common_pixels_angle_diff} 
quantitatively compares the orientation angles of common fibril pixels calculated by the fibril-fitting algorithms 
used in FibrilNet and the tool of \citet{Jing_2011}
based on the test image at 0.0 \AA \hspace*{+0.015cm} from the H$\alpha$ line center 6563 \AA \hspace*{+0.015cm}
with a 70\arcsec \hspace*{+0.015cm}  circular FOV collected in AR 12665 on 2017 July 13 20:15:58 UT.
Figure \ref{fig: common_pixels_angle_diff}(A) shows the 
2D histogram
of the orientation angles of common fibril pixels produced by the two tools
where the $x$-axis ($y$-axis, respectively) represents
the orientation angles calculated by FibrilNet (the tool of \citet{Jing_2011}, respectively).
The 2D histogram
is computed by grouping common fibril pixels whose orientation angles are 
specified by their $x$ and $y$ coordinates into bins, and counting the common fibril pixels in a bin to compute
the color of the tile representing the bin.
The width of each bin equals 2 degrees.
It can be seen from Figure \ref{fig: common_pixels_angle_diff}(A) that the orientation angles of common fibril pixels
calculated by the two tools mostly agree with each other, 
which is consistent with the findings shown in Figure \ref{fig: orientation}.
Figure \ref{fig: common_pixels_angle_diff}(B) 
shows differences of the orientation angles of common fibril pixels
produced by the two tools.
It can be seen from Figure \ref{fig: common_pixels_angle_diff}(B)
that most of the common fibril pixels
have very small orientation angle differences,  displayed by purple color. 
For the common fibril pixels with large orientation angle differences, 
the orientation angles calculated by the quadratic function used in the tool of \citet{Jing_2011} are often incorrect
(see, for example, 
the fibrils highlighted by the small red circles
in Figure \ref{fig: common_pixels_angle_diff}(B) and Figure \ref{fig: orientation}).

\subsection{Application of FibrilNet to Other Data}
\label{sec:application}

In this series of experiments, we applied FibrilNet to other types of test images, 
including (i) a full-disk image from the Global Oscillation Network Group \citep[GONG;][]{1996Sci...272.1284H, pub.1132138000}
at the National Solar Observatory (NSO), 
(ii) a full-disk image from the Kanzelhöhe Solar Observatory \citep[KSO;][]{1999ASPC..184..314O, 2008CEAB...32....1O}, 
(iii) high-resolution superpenumbral fibrils from BBSO \citep{Jing_2019}, 
and (iv) two high-resolution quiet Sun regions from BBSO. 
The GONG full-disk H$\alpha$ LH (Learmonth Reduced H$\alpha$) data in (i) was collected on 2015 September 28 00:01:34 UT. 
The KSO full-disk H$\alpha$ Fi (Full-disk raw image) data in (ii) was collected on 2015 September 14 09:14:20 UT. 
The GONG and KSO full-disk images have relatively low resolution.
The BBSO superpenumbra of sunspots in (iii) was collected at H$\alpha$ $-$0.6 \AA \hspace*{+0.015cm} 
from AR 12661 (501E, 95N)  on 2017 June 4 19:08:44 UT. 
The two BBSO quiet-Sun regions in (iv) were collected on 2018 July 29 16:33:12 UT and 2020 June 10 16:10:25 UT 
at H$\alpha$ $-$0.6 \AA \hspace*{+0.015cm} from (604E, 125S) 
and H$\alpha$ 0.0 \AA \hspace*{+0.015cm} from (283E, 789N), respectively. 
The FibrilNet tool was trained using the same 241 H$\alpha$ line center images described in Section \ref{sec:observational data}. 
Here we present results without uncertainty maps.
Results with uncertainty maps can be generated similarly as done in Section \ref{Tracing Halpha results}.

\begin{figure}
\epsscale{1.15}
\plotone{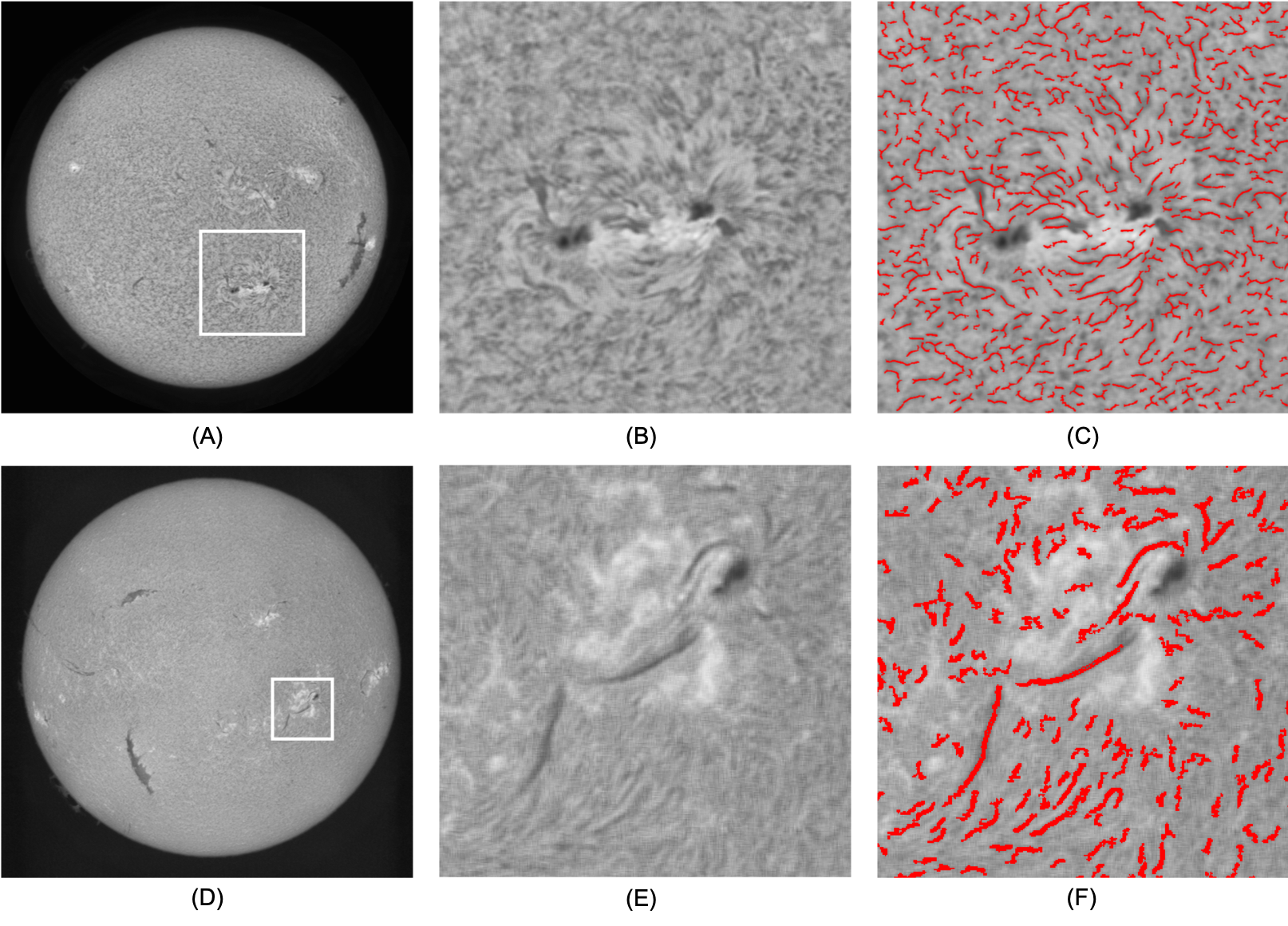}
\caption{Fibrils (red curves) predicted by FibrilNet on the 
GONG and KSO full-disk H$\alpha$ images 
collected on 2015 September 28 00:01:34 UT and 2015 September 14 09:14:20 UT, respectively. 
(A) The GONG full-disk H$\alpha$ image.
(B) The enlarged view of the region highlighted by the white box in (A).
(C) Fibrils predicted by FibrilNet on the image in (B).
(D) The KSO full-disk H$\alpha$ image.
(E) The enlarged view of the region highlighted by the white box in (D).
(F) Fibrils predicted by FibrilNet on the image in (E).
\label{fig: results_GONG_KSO}} 
\end{figure}

Figure \ref{fig: results_GONG_KSO} shows fibrils (red curves) 
predicted by FibrilNet on the
GONG and KSO test images. 
Figure \ref{fig: results_GONG_KSO}(A) presents the GONG full-disk H$\alpha$ image.
Figure \ref{fig: results_GONG_KSO}(B) shows the enlarged view of the region 
highlighted by the white box in Figure \ref{fig: results_GONG_KSO}(A).
In Figure \ref{fig: results_GONG_KSO}(C), we see that FibrilNet detects many fibrils on the GONG image.
Figure \ref{fig: results_GONG_KSO}(D) presents the KSO full-disk H$\alpha$ image.
Figure \ref{fig: results_GONG_KSO}(E) shows the enlarged view of the region highlighted 
by the white box in Figure \ref{fig: results_GONG_KSO}(D).
Figure \ref{fig: results_GONG_KSO}(F) clearly demonstrates that FibrilNet detects the threads of filaments and fibrils 
on the KSO image.

\begin{figure}
\epsscale{1.15}
\plotone{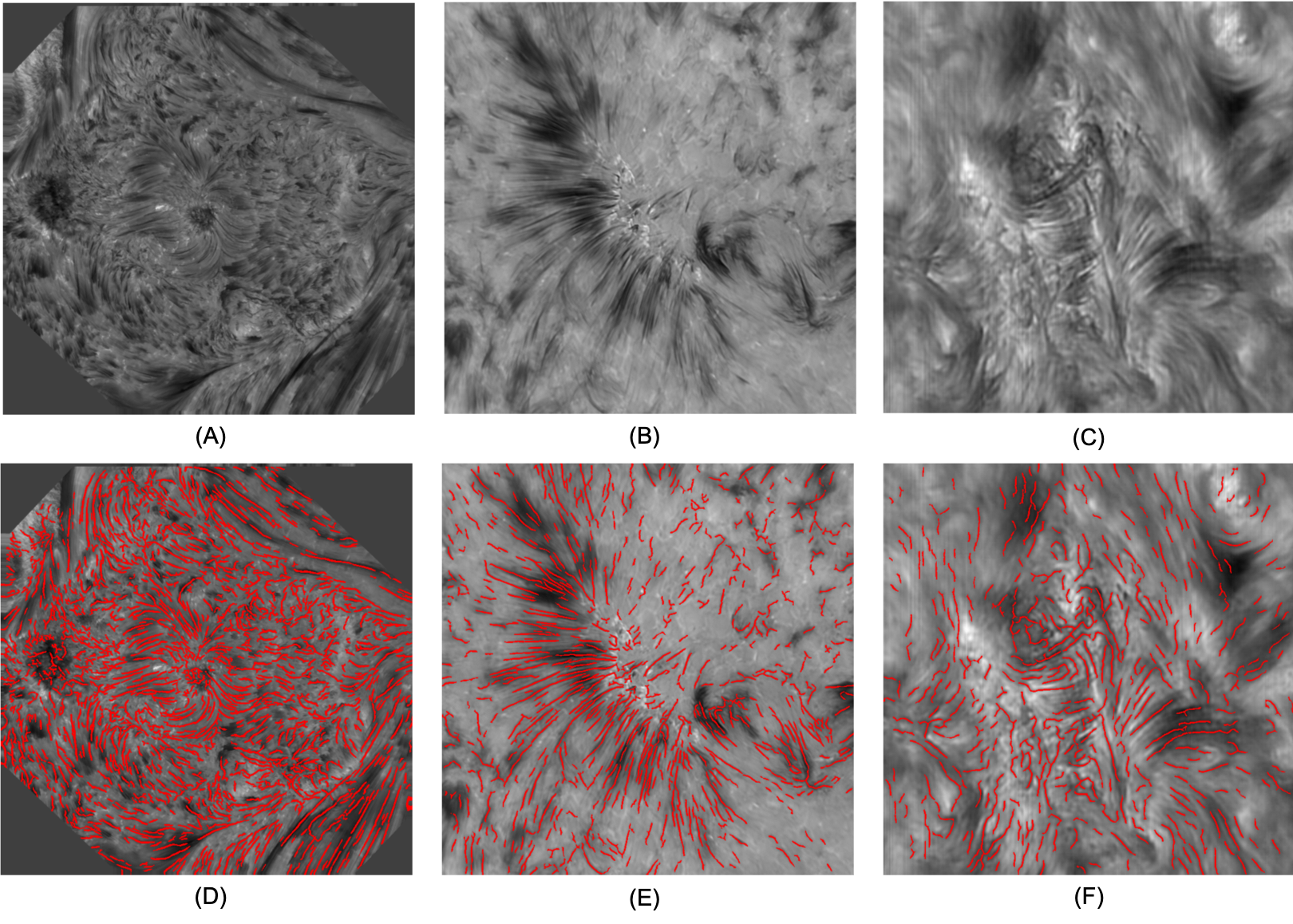}
\caption{Fibrils (red curves) predicted by FibrilNet on additional high-resolution BBSO test H$\alpha$ images.
(A) The BBSO superpenumbra of sunspots image collected at H$\alpha$ $-$0.6 \AA \hspace*{+0.015cm} 
from AR 12661 (501E, 95N)  on 2017 June 4 19:08:44 UT.
(B) The BBSO quiet-Sun image collected at H$\alpha$ $-$0.6 \AA \hspace*{+0.015cm} from (604E, 125S) on 2018 July 29 16:33:12 UT.
(C) The BBSO quiet-Sun image collected at H$\alpha$ 0.0 \AA \hspace*{+0.015cm} from (283E, 789N) on 2020 June 10 16:10:25 UT.
(D) Fibrils predicted by FibrilNet on the image in (A).
(E) Fibrils predicted by FibrilNet on the image in (B).
(F) Fibrils predicted by FibrilNet on the image in (C).
\label{fig: results_BBSO}}
\end{figure}

Figure \ref{fig: results_BBSO} presents fibril prediction results 
on the BBSO high-resolution test H$\alpha$ images. 
Figure \ref{fig: results_BBSO}(A) shows the BBSO superpenumbra of sunspots image used in the study.
It can be seen that there are superpernumbral fibrils around the sunspot in the center of the image.
Figure \ref{fig: results_BBSO}(D) shows the predicted superpenumbral fibrils (red curves) 
produced by FibrilNet on the image in Figure \ref{fig: results_BBSO}(A). 
We see in Figure \ref{fig: results_BBSO}(D) that
FibrilNet can distinguish the superpenumbral fibrils from the clusters of spicules nearby. 
Figures \ref{fig: results_BBSO}(B) and \ref{fig: results_BBSO}(C) present the two BBSO quiet-Sun regions. 
Figures \ref{fig: results_BBSO}(E) and \ref{fig: results_BBSO}(F) show the predicted mottles in the quiet-Sun rosette structures  
in Figures \ref{fig: results_BBSO}(B) and \ref{fig: results_BBSO}(C), respectively. 
These high-resolution H$\alpha$ images clearly demonstrate the good fibril prediction capability of our tool.

\section{Discussion and Conclusions} 
\label{sec:conclusion}
We develop a Bayesian deep learning method, FibrilNet, for tracing chromospheric fibrils in H$\alpha$ images of solar observations. 
We apply the FibrilNet tool to high-resolution H$\alpha$ images from an active region (AR 12665)
collected by BBSO/GST on July 13, 2017. 
The tool performs well on these high-resolution H$\alpha$ images, predicting fibrils with uncertainty quantification 
and determining the orientations of the predicted fibrils.
We further apply FibrilNet to full-disk H$\alpha$ images from other solar observatories 
and additional high-resolution H$\alpha$ images collected by BBSO/GST,
demonstrating the tool's usability in diverse datasets.

Our main results are summarized as follows:
\begin{quote}
1. The encoder-decoder convolutional neural network
(i.e., the Bayesian deep learning model) used in FibrilNet, as illustrated in Figure \ref{fig:fibrilNet},
is an enhancement of two deep learning models, namely U-Net \citep{unet_nature},
based on which our SolarUnet \citep{Jiang_2020} for magnetic tracking was developed,
and SegNet \citep{7803544}.
FibrilNet predicts fibrils on a test H$\alpha$ image through image segmentation (i.e.,
predicting each pixel in the test H$\alpha$ image to be black indicating the pixel is part of a fibril or white indicating the pixel is not part of a fibril).
In computer vision and image processing, U-Net and SegNet are two of the best image segmentation models.
By combining these two models, FibrilNet produces good image segmentation (i.e., fibril prediction) results,
as described in Section \ref{sec:experiment}.

2. The training dataset used in this study comprises 241 high-resolution H$\alpha$ line center images
in AR 12665 collected by BBSO/GST from 20:16:32 UT to 22:41:30 UT on 2017 July 13.
After FibrilNet is trained on this dataset, we apply the trained model to predict fibrils on five high-resolution test H$\alpha$ images
from the same active region (AR 12665) collected by BBSO/GST on 2017 July 13 20:15:58 UT
as described in Section \ref{Tracing Halpha results}, as well as
an additional five test H$\alpha$ images including 
two full-disk H$\alpha$ images from GONG/KSO and three other high-resolution H$\alpha$ images collected by BBSO/GST
as described in Section \ref{sec:application}.
Our experimental results show that the Bayesian deep learning model employed by FibrilNet performs well
not only on the five high-resolution test H$\alpha$ images from AR 12665 that are not seen during training, 
but also on the additional five test H$\alpha$ images.
No further training is needed for FibrilNet to predict fibrils in the additional five test H$\alpha$ images.
This is achieved by the generalization and inference capabilities of the deep learning model used by FibrilNet.
On the other hand, the threshold-based tool in \citet{Jing_2011}
is tailored for the high-resolution H$\alpha$  images collected by BBSO/GST. 
When applying the threshold-based tool in \citet{Jing_2011} to the GONG full-disk H$\alpha$ image 
in Figure \ref{fig: results_GONG_KSO},
the threshold-based tool performs poorly, missing many fibrils on the GONG H$\alpha$ image.

3. FibrilNet obtains training data from the threshold-based tool in \citet{Jing_2011}
where the training dataset contains 241 high-resolution H$\alpha$ line center images from AR 12665 
collected by BBSO/GST as described in item 2 above. 
When applying FibrilNet and the threshold-based tool to the five high-resolution test H$\alpha$ images 
from the same active region (AR 12665) collected by BBSO/GST, 
the two tools agree well on the detected fibrils as described in Sections \ref{Tracing Halpha results} and \ref{sec:quantitative}. 
This demonstrates the good learning capability of FibrilNet.
When predicting fibrils on a test  H$\alpha$ image, FibrilNet uses an uncertainty quantification technique
(more precisely a Monte Carlo sampling technique) to process the test H$\alpha$ image $T$ times where
$T = 50$ as described in Section \ref{sec:implementation}.
Unlike FibrilNet, which employs deep learning, the tool in \citet{Jing_2011} used a 
threshold-based algorithm, rather than machine learning, for image segmentation to detect fibrils 
on the test H$\alpha$ image.
It takes several seconds for the threshold-based tool to process the test H$\alpha$ image.
When the uncertainty quantification technique is turned off (i.e., $T$ is set to 1),
FibrilNet is ten times faster than the threshold-based tool in \citet{Jing_2011} due to the fact that 
FibrilNet detects fibrils through making predictions, while the two tools produce similar results.
When the uncertainty quantification technique is turned on (i.e., $T$ is set to 50),
FibrilNet is as fast as the threshold-based tool while producing uncertainty maps
that not only provide a quantitative way to measure the confidence on each detected fibril, 
but also help identify fibril structures that are not detected by the threshold-based tool
(i.e., that do not exist in the training data)
but are inferred through machine learning
as described in Section \ref{Tracing Halpha results}.
It is worth noting that the main source of uncertainty comes from the data rather than our deep learning model.
Uncertainty values are higher in the noisy regions of the test H$\alpha$ image.
Furthermore, the ends of a predicted fibril are often associated with higher uncertainty,
due to the ambiguity surrounding the transition from the fibril body to the non-fibril background area.
To the best of our knowledge, FibrilNet is the first tool capable of predicting fibrils with uncertainty quantification.

4. We conducted additional experiments to evaluate the effectiveness of 
the data augmentation technique used for training FibrilNet as described in Section \ref{sec:overview}.
Our experimental results show that, without the data augmentation technique, 
the performance of FibrilNet degrades,
particularly when the tool is applied to the GONG and KSO full-disk H$\alpha$ images
in Figure \ref{fig: results_GONG_KSO}.
This happens because the data augmentation technique can increase the generalization and inference capabilities 
of the Bayesian deep learning model used by FibrilNet.  
Our training dataset comprises
241 H$\alpha$ line center images from AR 12665 collected on July 13, 2017 as described in Section \ref{sec:observational data}.
We also performed experiments where we split the training dataset into
two parts based on image quality.
The first part contained 12  H$\alpha$ line center images with slightly lower quality.
The second part contained the remaining 229 H$\alpha$ line center images with higher quality.
Since the first part contained too few H$\alpha$ images, we expanded it by including 12 lower-quality images
from the other four wavelength positions in AR 12665 studied here, yielding a total of 60 lower-quality H$\alpha$ images. 
Our experimental results show that the deep learning models trained by all 241 
H$\alpha$ line center images and by the 229 higher-quality H$\alpha$ line center images
produce similar results.
On the other hand, the performance of the deep learning model trained by the 60 lower-quality H$\alpha$ images degrades,
and becomes even worse in the absence of data augmentation, particularly when the model is applied to the  
GONG and KSO full-disk H$\alpha$ images.

5. To further understand the behavior of FibrilNet, we trained the tool using all $241 \times 5 = 1205$ high-resolution H$\alpha$  images 
from all five wavelength positions in AR 12665 studied here
and applied the trained tool to the same test images described in Section \ref{sec:experiment}.
The results obtained are similar to those presented here, 
indicating our tool works equally well even with fewer training images. 		
When the tool is trained by a much smaller dataset such as one with less than
100 H$\alpha$ line center images from AR 12665 collected on July 13, 2017,
the tool still performs well on the high-resolution test H$\alpha$ images described in Section \ref{sec:experiment},
but finds fragmented filaments and fibrils, rather than long, complete filaments and fibrils, on the KSO full-disk H$\alpha$ image
in Figure \ref{fig: results_GONG_KSO},
even when the tool is trained by the data augmentation technique with higher-quality training images.

6. As mentioned above, the Bayesian deep learning model in FibrilNet 
performs image segmentation to predict fibrils with uncertainty quantification.
On the other hand, the fibril-fitting algorithm in FibrilNet uses a polynomial regression function with varying degrees
to calculate the orientation angles of the predicted fibrils.
This polynomial regression model produces more accurate and smoother 
fibril orientation angles than the quadratic function used by the tool in \citet{Jing_2011}
as described in Sections \ref{Tracing Halpha results} and \ref{sec:quantitative}.
However, if we replace the polynomial regression model by the quadratic function in FibrilNet,
the two tools would produce the same orientation angles on common fibril pixels detected by the tools.
\end{quote}

We conclude that FibrilNet is an effective and alternative method for fibril tracing.  
It is expected that this tool will be a useful utility for processing observations from diverse instruments 
including BBSO/GST and the new DKIST (Daniel K. Inouye Solar Telescope).

We thank the referee and Scientific Editor for very helpful and thoughtful comments.
We also thank the BBSO/GST team for providing the data used in this study. 
The BBSO operation is supported by the New Jersey Institute of Technology
and U.S. NSF grant AGS-1821294. 
The GST operation is partly supported by the Korea Astronomy and Space Science Institute, 
the Seoul National University, and the Key Laboratory of Solar Activities of the Chinese Academy of Sciences (CAS) 
and the Operation, Maintenance and Upgrading Fund of CAS for Astronomical Telescopes and Facility Instruments. 
This work was supported by U.S. NSF grants AGS-1927578 and AGS-1954737.
C.L. and H.W. acknowledge the support of NASA under grants 80NSSC20K1282, 
80NSSC18K0673, and 80NSSC18K1705.

\facilities{Big Bear Solar Observatory, National Solar Observatory, Kanzelhöhe Solar Observatory.}

\appendix
\renewcommand\thefigure{A\arabic{figure}}   
\setcounter{figure}{0} 

\begin{figure}
\epsscale{1.05}
\plotone{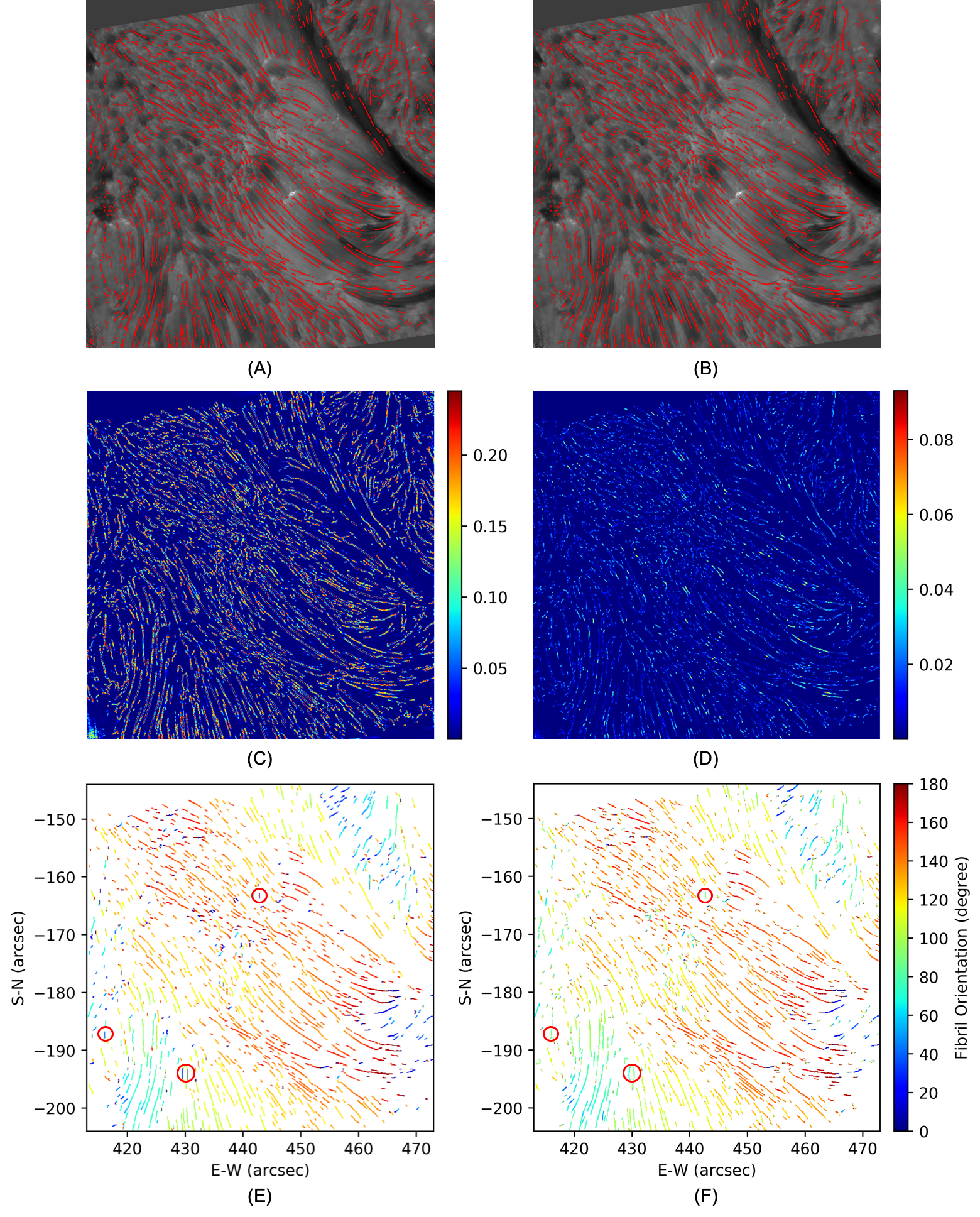}
\caption{Fibril tracing results on the test image at +0.4 \AA \hspace*{+0.015cm} 
from the H$\alpha$ line center 6563 \AA \hspace*{+0.015cm} 
with a 70\arcsec \hspace*{+0.015cm}  circular FOV collected in  AR 12665 on 2017 July 13 20:15:58 UT
where training data were 241 H$\alpha$ line center images taken from the same AR 
between 20:16:32 UT and 22:41:30 UT on the same day.
(A) Fibrils on the test H$\alpha$ image detected by the tool in \citet{Jing_2011}.
(B) Fibrils on the test H$\alpha$ image predicted by FibrilNet.
(C) The aleatoric uncertainty (data uncertainty) map produced by FibrilNet.
(D) The epistemic uncertainty (model uncertainty) map produced by FibrilNet. 
(E) Fibril orientation angles calculated by the tool in \citet{Jing_2011}.
(F) Fibril orientation angles determined by FibrilNet.
Orientation angles of a number of fibrils,
some of which are highlighted by small red circles here,
are calculated wrongly by the tool in \citet{Jing_2011}, 
but correctly by FibrilNet.
\label{fig: results_har040}}
\end{figure}

\begin{figure}
\epsscale{1.05}
\plotone{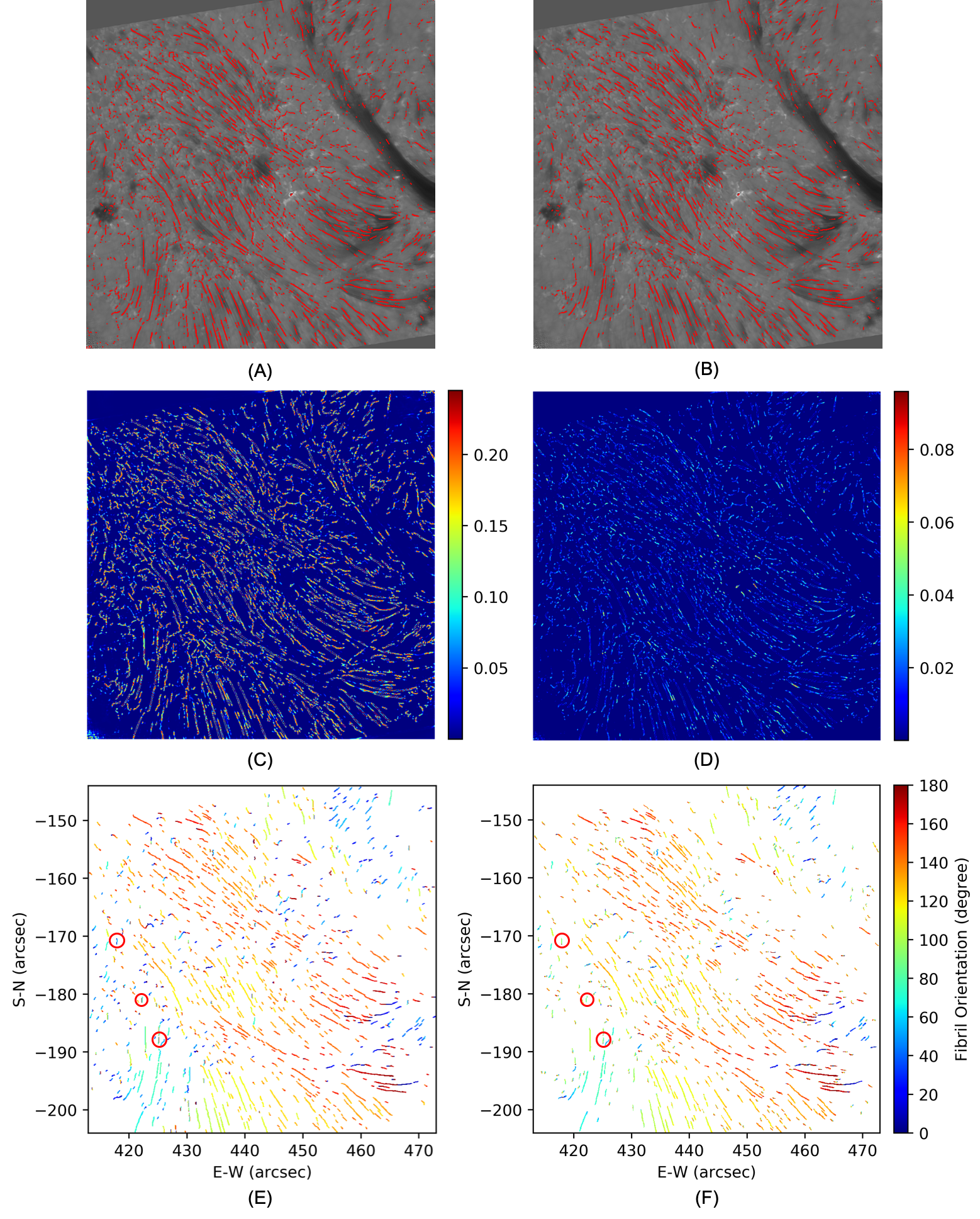}
\caption{Fibril tracing results on the test image at +0.6 \AA \hspace*{+0.015cm} 
from the H$\alpha$ line center 6563 \AA \hspace*{+0.015cm} 
with a 70\arcsec \hspace*{+0.015cm}  circular FOV collected in AR 12665 on 2017 July 13 20:15:58 UT
where training data were 241 H$\alpha$ line center images taken from the same AR 
between 20:16:32 UT and 22:41:30 UT on the same day.
(A) Fibrils on the test H$\alpha$ image detected by the tool in \citet{Jing_2011}.
(B) Fibrils on the test H$\alpha$ image predicted by FibrilNet.
(C) The aleatoric uncertainty (data uncertainty) map produced by FibrilNet.
(D) The epistemic uncertainty (model uncertainty) map produced by FibrilNet. 
(E) Fibril orientation angles calculated by the tool in \citet{Jing_2011}.
(F) Fibril orientation angles determined by FibrilNet.
Orientation angles of a number of fibrils,
some of which are highlighted by small red circles here,
are calculated wrongly by the tool in \citet{Jing_2011}, 
but correctly by FibrilNet.
\label{fig: results_har060}}
\end{figure}

\begin{figure}
\epsscale{1.05}
\plotone{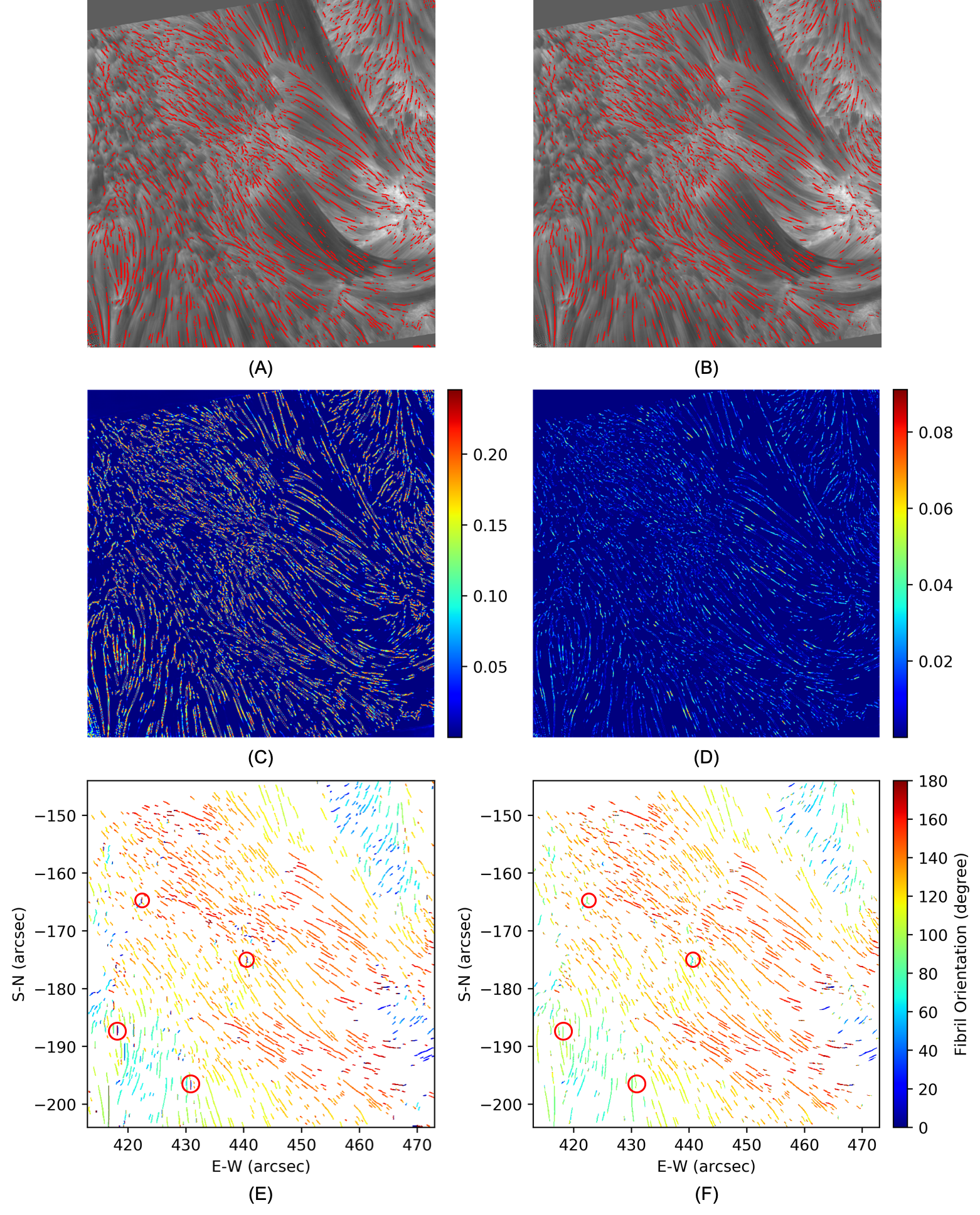}
\caption{Fibril tracing results on the test image at $-$0.4 \AA \hspace*{+0.015cm} 
from the H$\alpha$ line center 6563 \AA \hspace*{+0.015cm} 
with a 70\arcsec \hspace*{+0.015cm}  circular FOV collected in AR 12665 on 2017 July 13 20:15:58 UT
where training data were 241 H$\alpha$ line center images taken from the same AR 
between 20:16:32 UT and 22:41:30 UT on the same day.
(A) Fibrils on the test H$\alpha$ image detected by the tool in \citet{Jing_2011}.
(B) Fibrils on the test H$\alpha$ image predicted by FibrilNet.
(C) The aleatoric uncertainty (data uncertainty) map produced by FibrilNet.
(D) The epistemic uncertainty (model uncertainty) map produced by FibrilNet. 
(E) Fibril orientation angles calculated by the tool in \citet{Jing_2011}.
(F) Fibril orientation angles determined by FibrilNet.
Orientation angles of a number of fibrils,
some of which are highlighted by small red circles here,
are calculated wrongly by the tool in \citet{Jing_2011}, 
but correctly by FibrilNet.
\label{fig: results_hab040}}
\end{figure}

\begin{figure}
\epsscale{1.05}
\plotone{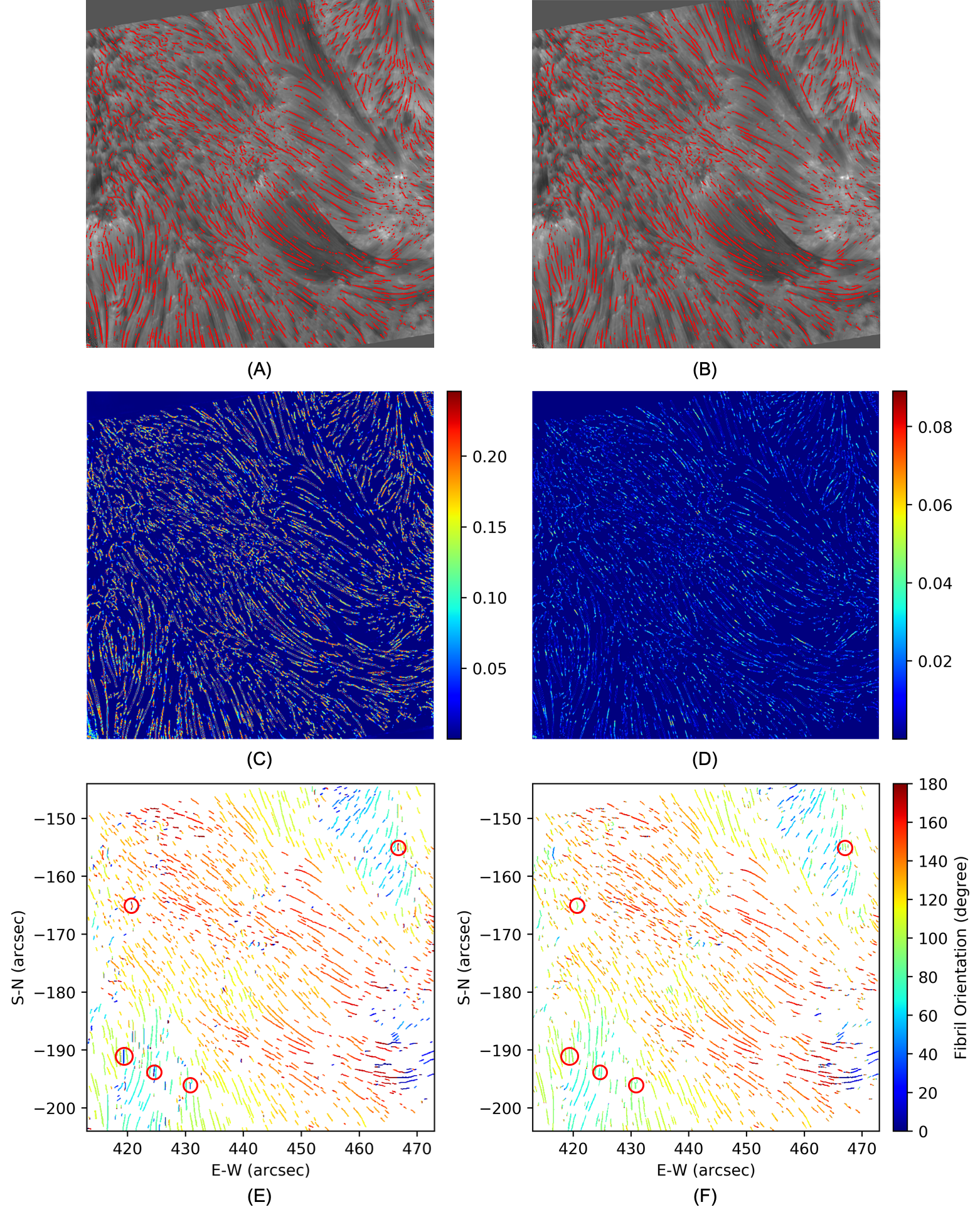}
\caption{Fibril tracing results on the test image at $-$0.6 \AA \hspace*{+0.015cm} 
from the H$\alpha$ line center 6563 \AA \hspace*{+0.015cm} 
with a 70\arcsec \hspace*{+0.015cm}  circular FOV collected in AR 12665 on 2017 July 13 20:15:58 UT
where training data were 241 H$\alpha$ line center images taken from the same AR 
between 20:16:32 UT and 22:41:30 UT on the same day.
(A) Fibrils on the test H$\alpha$ image detected by the tool in \citet{Jing_2011}.
(B) Fibrils on the test H$\alpha$ image predicted by FibrilNet.
(C) The aleatoric uncertainty (data uncertainty) map produced by FibrilNet.
(D) The epistemic uncertainty (model uncertainty) map produced by FibrilNet. 
(E) Fibril orientation angles calculated by the tool in \citet{Jing_2011}.
(F) Fibril orientation angles determined by FibrilNet.
Orientation angles of a number of fibrils,
some of which are highlighted by small red circles here,
are calculated wrongly by the tool in \citet{Jing_2011}, 
but correctly by FibrilNet.
\label{fig: results_hab060}}
\end{figure}

Figure \ref{fig: results_har040} (Figure \ref{fig: results_har060}, Figure \ref{fig: results_hab040}, Figure \ref{fig: results_hab060}, respectively) 
compares fibril tracing results and fibril orientations obtained by FibrilNet and the tool in \citet{Jing_2011} on the
test image at +0.4 \AA  \hspace*{+0.015cm} (+0.6 \AA, $-$0.4 \AA, $-$0.6 \AA, respectively) 
from the H$\alpha$ line center 6563 \AA \hspace*{+0.015cm} with a 70\arcsec \hspace*{+0.015cm}  
circular FOV collected in AR 12665 on 2017 July 13 20:15:58 UT
where training data were 241 H$\alpha$ line center images taken from the same AR between 20:16:32 UT and 22:41:30 UT on the same day.

\bibliographystyle{aasjournal}
\bibliography{reference}

\begin{thebibliography}{}
\expandafter\ifx\csname natexlab\endcsname\relax\def\natexlab#1{#1}\fi
\providecommand{\url}[1]{\href{#1}{#1}}
\providecommand{\dodoi}[1]{doi:~\href{http://doi.org/#1}{\nolinkurl{#1}}}
\providecommand{\doeprint}[1]{\href{http://ascl.net/#1}{\nolinkurl{http://ascl.net/#1}}}
\providecommand{\doarXiv}[1]{\href{https://arxiv.org/abs/#1}{\nolinkurl{https://arxiv.org/abs/#1}}}

\bibitem[{{Aschwanden} {et~al.}(2016){Aschwanden}, {Reardon}, \&
  {Jess}}]{2016ApJ...826...61A}
{Aschwanden}, M.~J., {Reardon}, K., \& {Jess}, D.~B. 2016, \apj, 826, 61,
  \dodoi{10.3847/0004-637X/826/1/61}

\bibitem[{{Asensio Ramos} {et~al.}(2017){Asensio Ramos}, {de la Cruz
  Rodr{\'{\i}}guez}, {Mart{\'{\i}}nez Gonz{\'a}lez}, \&
  {Socas-Navarro}}]{2017A&A...599A.133A}
{Asensio Ramos}, A., {de la Cruz Rodr{\'{\i}}guez}, J., {Mart{\'{\i}}nez
  Gonz{\'a}lez}, M.~J., \& {Socas-Navarro}, H. 2017, \aap, 599, A133,
  \dodoi{10.1051/0004-6361/201629755}

\bibitem[{{Badrinarayanan} {et~al.}(2017){Badrinarayanan}, {Kendall}, \&
  {Cipolla}}]{7803544}
{Badrinarayanan}, V., {Kendall}, A., \& {Cipolla}, R. 2017, IEEE Transactions
  on Pattern Analysis and Machine Intelligence, 39, 2481,
  \dodoi{10.1109/TPAMI.2016.2644615}

\bibitem[{Bishop(2006)}]{10.5555/1162264}
Bishop, C.~M. 2006, Pattern Recognition and Machine Learning (Information
  Science and Statistics) (Berlin, Heidelberg: Springer-Verlag).
\newblock \url{https://dl.acm.org/doi/book/10.5555/1162264}

\bibitem[{Blei {et~al.}(2017)Blei, Kucukelbir, \&
  McAuliffe}]{doi:10.1080/01621459.2017.1285773}
Blei, D.~M., Kucukelbir, A., \& McAuliffe, J.~D. 2017, Journal of the American
  Statistical Association, 112, 859, \dodoi{10.1080/01621459.2017.1285773}

\bibitem[{{Cao} {et~al.}(2010){Cao}, {Gorceix}, {Coulter}, {Ahn}, {Rimmele}, \&
  {Goode}}]{2010AN....331..636C}
{Cao}, W., {Gorceix}, N., {Coulter}, R., {et~al.} 2010, Astronomische
  Nachrichten, 331, 636, \dodoi{10.1002/asna.201011390}

\bibitem[{Denker \& LeCun(1990)}]{10.5555/2986766.2986882}
Denker, J.~S., \& LeCun, Y. 1990, in Proceedings of the 3rd International
  Conference on Neural Information Processing Systems, NIPS'90 (San Francisco,
  CA, USA: Morgan Kaufmann Publishers Inc.), 853–859.
\newblock \url{https://dl.acm.org/doi/10.5555/2986766.2986882}

\bibitem[{Falk {et~al.}(2019)Falk, Mai, Bensch, {\c C}i{\c c}ek, Abdulkadir,
  Marrakchi, B{\"o}hm, Deubner, J{\"a}ckel, Seiwald, Dovzhenko, Tietz,
  Dal~Bosco, Walsh, Saltukoglu, Tay, Prinz, Palme, Simons, Diester, Brox, \&
  Ronneberger}]{unet_nature}
Falk, T., Mai, D., Bensch, R., {et~al.} 2019, Nature Methods, 16, 67,
  \dodoi{10.1038/s41592-018-0261-2}

\bibitem[{{Fleishman} {et~al.}(2019){Fleishman}, {Mysh'yakov}, {Stupishin},
  {Loukitcheva}, \& {Anfinogentov}}]{2019ApJ...870..101F}
{Fleishman}, G., {Mysh'yakov}, I., {Stupishin}, A., {Loukitcheva}, M., \&
  {Anfinogentov}, S. 2019, \apj, 870, 101, \dodoi{10.3847/1538-4357/aaf384}

\bibitem[{Fossum \& Carlsson(2006)}]{Fossum_2006}
Fossum, A., \& Carlsson, M. 2006, \apj, 646, 579, \dodoi{10.1086/504887}

\bibitem[{{Foukal}(1971{\natexlab{a}})}]{1971SoPh...20..298F}
{Foukal}, P. 1971{\natexlab{a}}, \solphys, 20, 298, \dodoi{10.1007/BF00159759}

\bibitem[{{Foukal}(1971{\natexlab{b}})}]{1971SoPh...19...59F}
---. 1971{\natexlab{b}}, \solphys, 19, 59, \dodoi{10.1007/BF00148824}

\bibitem[{{Gafeira} {et~al.}(2017){Gafeira}, {Lagg}, {Solanki}, {Jafarzadeh},
  {van Noort}, {Barthol}, {Blanco Rodr{\'{\i}}guez}, {del Toro Iniesta},
  {Gandorfer}, {Gizon}, {Hirzberger}, {Kn{\"o}lker}, {Orozco Su{\'a}rez},
  {Riethm{\"u}ller}, \& {Schmidt}}]{2017ApJS..229....6G}
{Gafeira}, R., {Lagg}, A., {Solanki}, S.~K., {et~al.} 2017, \apjs, 229, 6,
  \dodoi{10.3847/1538-4365/229/1/6}

\bibitem[{Gal \& Ghahramani(2016)}]{10.5555/3045390.3045502}
Gal, Y., \& Ghahramani, Z. 2016, in Proceedings of the 33rd International
  Conference on International Conference on Machine Learning - Volume 48,
  ICML'16 (JMLR.org), 1050–1059.
\newblock \url{https://dl.acm.org/doi/10.5555/3045390.3045502}

\bibitem[{{Goode} \& {Cao}(2012)}]{2012SPIE.8444E..03G}
{Goode}, P.~R., \& {Cao}, W. 2012, in Ground-based and Airborne Telescopes IV,
  ed. L.~M. Stepp, R.~Gilmozzi, \& H.~J. Hall, Vol. 8444, International Society
  for Optics and Photonics (SPIE), 1--8, \dodoi{10.1117/12.925494}

\bibitem[{{Goode} {et~al.}(2010){Goode}, {Yurchyshyn}, {Cao}, {Abramenko},
  {Andic}, {Ahn}, \& {Chae}}]{2010ApJ...714L..31G}
{Goode}, P.~R., {Yurchyshyn}, V., {Cao}, W., {et~al.} 2010, \apjl, 714, L31,
  \dodoi{10.1088/2041-8205/714/1/L31}

\bibitem[{Goodfellow {et~al.}(2016)Goodfellow, Bengio, \&
  Courville}]{Goodfellow-et-al-2016}
Goodfellow, I., Bengio, Y., \& Courville, A. 2016, Deep Learning (MIT Press).
\newblock \url{http://www.deeplearningbook.org}

\bibitem[{Graves(2011)}]{10.5555/2986459.2986721}
Graves, A. 2011, in Advances in Neural Information Processing Systems 24: 25th
  Annual Conference on Neural Information Processing Systems 2011. Proceedings
  of a meeting held 12-14 December 2011, Granada, Spain, ed. J.~Shawe{-}Taylor,
  R.~S. Zemel, P.~L. Bartlett, F.~C.~N. Pereira, \& K.~Q. Weinberger,
  2348--2356.
\newblock
  \url{http://papers.nips.cc/paper/4329-practical-variational-inference-for-neural-networks}

\bibitem[{{Gruet} {et~al.}(2018){Gruet}, {Chandorkar}, {Sicard}, \&
  {Camporeale}}]{GCS-2018}
{Gruet}, M.~A., {Chandorkar}, M., {Sicard}, A., \& {Camporeale}, E. 2018, Space
  Weather, 16, 1882, \dodoi{10.1029/2018SW001898}

\bibitem[{{Harvey} {et~al.}(1996){Harvey}, {Hill}, {Hubbard}, {Kennedy},
  {Leibacher}, {Pintar}, {Gilman}, {Noyes}, {Title}, {Toomre}, {Ulrich},
  {Bhatnagar}, {Kennewell}, {Marquette}, {Patron}, {Saa}, \&
  {Yasukawa}}]{1996Sci...272.1284H}
{Harvey}, J.~W., {Hill}, F., {Hubbard}, R.~P., {et~al.} 1996, Science, 272,
  1284, \dodoi{10.1126/science.272.5266.1284}

\bibitem[{He {et~al.}(2009)He, Chao, Suzuki, \&
  Wu}]{He:2009:FCL:1542560.1542851}
He, L., Chao, Y., Suzuki, K., \& Wu, K. 2009, Pattern Recogn., 42, 1977,
  \dodoi{10.1016/j.patcog.2008.10.013}

\bibitem[{{Heinzel} \& {Schmieder}(1994)}]{1994A&A...282..939H}
{Heinzel}, P., \& {Schmieder}, B. 1994, \aap, 282, 939

\bibitem[{Huertas-Company {et~al.}(2018)Huertas-Company, Primack, Dekel, Koo,
  Lapiner, Ceverino, Simons, Snyder, Bernardi, Chen,
  Dom{\'{\i}}nguez-S{\'{a}}nchez, Lee, Margalef-Bentabol, \&
  Tuccillo}]{Huertas_Company_2018}
Huertas-Company, M., Primack, J.~R., Dekel, A., {et~al.} 2018, \apj, 858, 114,
  \dodoi{10.3847/1538-4357/aabfed}

\bibitem[{{Jafarzadeh} {et~al.}(2017){Jafarzadeh}, {Solanki}, {Gafeira}, {van
  Noort}, {Barthol}, {Blanco Rodr{\'{\i}}guez}, {del Toro Iniesta}, { dorfer},
  {Gizon}, {Hirzberger}, {Kn{\"o}lker}, {Orozco Su{\'a}rez}, {Riethm{\"u}ller},
  \& {Schmidt}}]{2017ApJS..229....9J}
{Jafarzadeh}, S., {Solanki}, S.~K., {Gafeira}, R., {et~al.} 2017, \apjs, 229,
  9, \dodoi{10.3847/1538-4365/229/1/9}

\bibitem[{Jiang {et~al.}(2020)Jiang, Wang, Liu, Jing, Liu, Wang, \&
  Wang}]{Jiang_2020}
Jiang, H., Wang, J., Liu, C., {et~al.} 2020, \apjs, 250, 5,
  \dodoi{10.3847/1538-4365/aba4aa}

\bibitem[{Jing {et~al.}(2019)Jing, Li, Liu, Lee, Xu, Cao, \& Wang}]{Jing_2019}
Jing, J., Li, Q., Liu, C., {et~al.} 2019, \apj, 880, 143,
  \dodoi{10.3847/1538-4357/ab2b44}

\bibitem[{Jing {et~al.}(2011)Jing, Yuan, Reardon, Wiegelmann, Xu, \&
  Wang}]{Jing_2011}
Jing, J., Yuan, Y., Reardon, K., {et~al.} 2011, \apj, 739, 67,
  \dodoi{10.1088/0004-637x/739/2/67}

\bibitem[{Kendall \& Gal(2017)}]{10.5555/3295222.3295309}
Kendall, A., \& Gal, Y. 2017, in Proceedings of the 31st International
  Conference on Neural Information Processing Systems, NIPS'17 (Red Hook, NY,
  USA: Curran Associates Inc.), 5580–5590.
\newblock \url{https://dl.acm.org/doi/10.5555/3295222.3295309}

\bibitem[{Kim {et~al.}(2019)Kim, Park, Lee, Moon, Bae, Lim, Jang, Kim, Cho,
  Choi, \& Cho}]{Kim2019}
Kim, T., Park, E., Lee, H., {et~al.} 2019, Nature Astronomy, 3, 397,
  \dodoi{10.1038/s41550-019-0711-5}

\bibitem[{Kwon {et~al.}(2020)Kwon, Won, Kim, \& Paik}]{KWON2020106816}
Kwon, Y., Won, J.-H., Kim, B.~J., \& Paik, M.~C. 2020, Computational Statistics
  and Data Analysis, 142, 106816,
  \dodoi{https://doi.org/10.1016/j.csda.2019.106816}

\bibitem[{{Langangen} {et~al.}(2008){Langangen}, {De Pontieu}, {Carlsson},
  {Hansteen}, {Cauzzi}, \& {Reardon}}]{2008ApJ...679L.167L}
{Langangen}, {\O}., {De Pontieu}, B., {Carlsson}, M., {et~al.} 2008, \apjl,
  679, L167, \dodoi{10.1086/589442}

\bibitem[{LeCun {et~al.}(2015)LeCun, Bengio, \& Hinton}]{LeCun2015}
LeCun, Y., Bengio, Y., \& Hinton, G. 2015, Nature, 521, 436,
  \dodoi{10.1038/nature14539}

\bibitem[{Leenaarts {et~al.}(2015)Leenaarts, Carlsson, \& {Rouppe van der
  Voort}}]{Leenaarts_2015}
Leenaarts, J., Carlsson, M., \& {Rouppe van der Voort}, L. 2015, \apj, 802,
  136, \dodoi{10.1088/0004-637x/802/2/136}

\bibitem[{Leung \& Bovy(2018)}]{10.1093/mnras/sty3217}
Leung, H.~W., \& Bovy, J. 2018, Monthly Notices of the Royal Astronomical
  Society, 483, 3255, \dodoi{10.1093/mnras/sty3217}

\bibitem[{Lieu {et~al.}(2019)Lieu, Conversi, Altieri, \&
  Carry}]{10.1093/mnras/stz761}
Lieu, M., Conversi, L., Altieri, B., \& Carry, B. 2019, Monthly Notices of the
  Royal Astronomical Society, 485, 5831, \dodoi{10.1093/mnras/stz761}

\bibitem[{{Liu} {et~al.}(2019){Liu}, {Liu}, {Wang}, \&
  {Wang}}]{2019ApJ...877..121L}
{Liu}, H., {Liu}, C., {Wang}, J. T.~L., \& {Wang}, H. 2019, \apj, 877, 121,
  \dodoi{10.3847/1538-4357/ab1b3c}

\bibitem[{{Loughhead}(1968)}]{1968SoPh....5..489L}
{Loughhead}, R.~E. 1968, \solphys, 5, 489, \dodoi{10.1007/BF00147015}

\bibitem[{{Martin}(1998)}]{1998SoPh..182..107M}
{Martin}, S.~F. 1998, \solphys, 182, 107, \dodoi{10.1023/A:1005026814076}

\bibitem[{{Mooroogen} {et~al.}(2017){Mooroogen}, {Morton}, \&
  {Henriques}}]{2017A&A...607A..46M}
{Mooroogen}, K., {Morton}, R.~J., \& {Henriques}, V. 2017, \aap, 607, A46,
  \dodoi{10.1051/0004-6361/201730926}

\bibitem[{Ostertagová(2012)}]{OSTERTAGOVA2012500}
Ostertagová, E. 2012, Procedia Engineering, 48, 500 ,
  \dodoi{https://doi.org/10.1016/j.proeng.2012.09.545}

\bibitem[{{Otruba}(1999)}]{1999ASPC..184..314O}
{Otruba}, W. 1999, in Astronomical Society of the Pacific Conference Series,
  Vol. 184, Third Advances in Solar Physics Euroconference: Magnetic Fields and
  Oscillations, ed. B.~{Schmieder}, A.~{Hofmann}, \& J.~{Staude}, 314--318.
\newblock \url{https://ui.adsabs.harvard.edu/abs/1999ASPC..184..314O}

\bibitem[{{Otruba} {et~al.}(2008){Otruba}, {Freislich}, \&
  {Hanslmeier}}]{2008CEAB...32....1O}
{Otruba}, W., {Freislich}, H., \& {Hanslmeier}, A. 2008, Central European
  Astrophysical Bulletin, 32, 1.
\newblock \url{https://ui.adsabs.harvard.edu/abs/2008CEAB...32....1O}

\bibitem[{{Pikel'ner}(1971)}]{1971SoPh...20..286P}
{Pikel'ner}, S.~B. 1971, \solphys, 20, 286, \dodoi{10.1007/BF00159757}

\bibitem[{Plowman \& Berger(2020)}]{pub.1132138000}
Plowman, J.~E., \& Berger, T.~E. 2020, Solar Physics, 295, 143,
  \dodoi{10.1007/s11207-020-01682-4}

\bibitem[{Rand(1971)}]{10.2307/2284239}
Rand, W.~M. 1971, Journal of the American Statistical Association, 66, 846.
\newblock \url{http://www.jstor.org/stable/2284239}

\bibitem[{{Rouppe van der Voort} {et~al.}(2009){Rouppe van der Voort},
  {Leenaarts}, {de Pontieu}, {Carlsson}, \& {Vissers}}]{2009ApJ...705..272R}
{Rouppe van der Voort}, L., {Leenaarts}, J., {de Pontieu}, B., {Carlsson}, M.,
  \& {Vissers}, G. 2009, \apj, 705, 272, \dodoi{10.1088/0004-637X/705/1/272}

\bibitem[{{Schad}(2017)}]{2017SoPh..292..132S}
{Schad}, T. 2017, \solphys, 292, 132, \dodoi{10.1007/s11207-017-1153-9}

\bibitem[{{Shumko} {et~al.}(2014){Shumko}, {Gorceix}, {Choi}, {Kellerer},
  {Cao}, {Goode}, {Abramenko}, {Richards}, {Rimmele}, \&
  {Marino}}]{2014SPIE.9148E..35S}
{Shumko}, S., {Gorceix}, N., {Choi}, S., {et~al.} 2014, in Society of
  Photo-Optical Instrumentation Engineers (SPIE) Conference Series, Vol. 9148,
  Adaptive Optics Systems IV, ed. E.~{Marchetti}, L.~M. {Close}, \& J.-P.
  {Vran}, 914835, \dodoi{10.1117/12.2056731}

\bibitem[{Umbaugh(2010)}]{Umbaugh:2010:DIP:1951634}
Umbaugh, S.~E. 2010, Digital Image Processing and Analysis: Human and Computer
  Vision Applications with CVIPtools, Second Edition, 2nd edn. (Boca Raton, FL,
  USA: CRC Press, Inc.).
\newblock \url{https://dl.acm.org/citation.cfm?id=1951634}

\bibitem[{{Unnikrishnan} {et~al.}(2005){Unnikrishnan}, {Pantofaru}, \&
  {Hebert}}]{1565332}
{Unnikrishnan}, R., {Pantofaru}, C., \& {Hebert}, M. 2005, in 2005 IEEE
  Computer Society Conference on Computer Vision and Pattern Recognition
  (CVPR'05) - Workshops, 34--34, \dodoi{10.1109/CVPR.2005.390}

\bibitem[{{Varsik} {et~al.}(2014){Varsik}, {Plymate}, {Goode}, {Kosovichev},
  {Cao}, {Coulter}, {Ahn}, {Gorceix}, \& {Shumko}}]{2014SPIE.9147E..5DV}
{Varsik}, J., {Plymate}, C., {Goode}, P., {et~al.} 2014, in \procspie, Vol.
  9147, Ground-based and Airborne Instrumentation for Astronomy V, 91475D,
  \dodoi{10.1117/12.2056688}

\bibitem[{Wang {et~al.}(2000)Wang, Li, Denker, Lee, Wang, Goode, McAllister, \&
  Martin}]{Wang_2000}
Wang, J., Li, W., Denker, C., {et~al.} 2000, The Astrophysical Journal, 530,
  1071, \dodoi{10.1086/308377}

\bibitem[{{Wiegelmann} {et~al.}(2008){Wiegelmann}, {Thalmann}, {Schrijver}, {De
  Rosa}, \& {Metcalf}}]{2008SoPh..247..249W}
{Wiegelmann}, T., {Thalmann}, J.~K., {Schrijver}, C.~J., {De Rosa}, M.~L., \&
  {Metcalf}, T.~R. 2008, \solphys, 247, 249, \dodoi{10.1007/s11207-008-9130-y}

\bibitem[{{W{\"o}ger} {et~al.}(2008){W{\"o}ger}, {von der L{\"u}he}, \&
  {Reardon}}]{2008A&A...488..375W}
{W{\"o}ger}, F., {von der L{\"u}he}, O., \& {Reardon}, K. 2008, \aap, 488, 375,
  \dodoi{10.1051/0004-6361:200809894}

\bibitem[{Wu \& Boada(2019)}]{10.1093/mnras/stz333}
Wu, J.~F., \& Boada, S. 2019, Monthly Notices of the Royal Astronomical
  Society, 484, 4683, \dodoi{10.1093/mnras/stz333}

\bibitem[{Xiao \& Wang(2019)}]{DBLP:conf/aaai/XiaoW19}
Xiao, Y., \& Wang, W.~Y. 2019, in The Thirty-Third {AAAI} Conference on
  Artificial Intelligence, {AAAI} 2019, The Thirty-First Innovative
  Applications of Artificial Intelligence Conference, {IAAI} 2019, The Ninth
  {AAAI} Symposium on Educational Advances in Artificial Intelligence, {EAAI}
  2019, Honolulu, Hawaii, USA, January 27 - February 1, 2019 ({AAAI} Press),
  7322--7329, \dodoi{10.1609/aaai.v33i01.33017322}

\bibitem[{{Xu} {et~al.}(2020){Xu}, {Huang}, {Yuan}, {Deng}, \&
  {Jiang}}]{XHY-2020}
{Xu}, S.~B., {Huang}, S.~Y., {Yuan}, Z.~G., {Deng}, X.~H., \& {Jiang}, K. 2020,
  \apjs, 248, 14, \dodoi{10.3847/1538-4365/ab880e}

\end{thebibliography}

\end{document}